\documentclass[12pt,american,english]{article}
\usepackage[T1]{fontenc}
\usepackage[latin9]{inputenc}
\usepackage{float}
\usepackage{amsmath}
\usepackage{amssymb}
\usepackage{graphicx}

\makeatletter

\providecommand{\tabularnewline}{\\}

\newcommand{\lyxaddress}[1]{
\par {\raggedright #1
\vspace{1.4em}
\noindent\par}
}

\makeatother

\usepackage{babel}
\begin{document}

\title{Financial Market Modeling with Quantum Neural Networks}

\author{Carlos Pedro Gonçalves}

\maketitle

\lyxaddress{Instituto Superior de Ciências Sociais e Políticas (ISCSP) - University
of Lisbon, E-mail: cgoncalves@iscsp.ulisboa.pt}
\begin{abstract}
Econophysics has developed as a research field that applies the formalism
of Statistical Mechanics and Quantum Mechanics to address Economics
and Finance problems. The branch of Econophysics that applies of Quantum
Theory to Economics and Finance is called Quantum Econophysics. In
Finance, Quantum Econophysics' contributions have ranged from option
pricing to market dynamics modeling, behavioral finance and applications
of Game Theory, integrating the empirical finding, from human decision
analysis, that shows that nonlinear update rules in probabilities,
leading to non-additive decision weights, can be computationally approached
from quantum computation, with resulting quantum interference terms
explaining the non-additive probabilities. The current work draws
on these results to introduce new tools from Quantum Artificial Intelligence,
namely Quantum Artificial Neural Networks as a way to build and simulate
financial market models with adaptive selection of trading rules,
leading to turbulence and excess kurtosis in the returns distributions
for a wide range of parameters.\newline

\end{abstract}

\section{Introduction}

One of the major problems of financial modeling has been to address
complex financial returns dynamics, in particular, excess kurtosis
and volatility-related turbulence which lead to statistically significant
deviations from the Gaussian random walk model worked in traditional
Financial Theory (Arthur et al., 1997; Voit, 2001; Ilinsky, 2001;
Focardi and Fabozzi, 2004). A main contribution of Econophysics to
Finance has been to address this problem using the tools from Statistical
Mechanics and Quantum Mechanics, within the paradigmatic basis of
Systems Science and Complexity Sciences (Anderson et al., 1988; Arthur
et al., 1997; Voit, 2001; Ehrentreich, 2008).

Econophysics is currently a major research area that has combined
interdisciplinarily Finance and Economics, Complex Systems Science,
Statistical Mechanics, Quantum Mechanics and Cognitive Science to
address notions and problems in Economics and Finance (Anderson et
al., 1988; Arthur et al., 1997; Voit, 2001; Brunn, 2006; Ehrentreich,
2008; Piotrowski and Sladkowski, 2001, 2002, 2008; Saptsin and Soloviev,
2009, 2011).

There are two major branches in Econophysics: Classical Econophysics
(based on Classical Mechanics) and Quantum Econophysics (based on
Quantum Mechanics). In Finance, Quantum Econophysics has been applied
to option pricing (Segal and Segal, 1998; Baaquie et al., 2000; Baaquie
and Marakani, 2001; Baaquie, 2004; Baaquie and Pan, 2011), financial
turbulence modeling (Gonçalves, 2011, 2013) and as an approach to
the formulation of financial theory, regarding price formation and
basic market relations (Piotrowski and Sladkowski, 2001, 2002, 2008;
Khrennikov, 2010; Haven and Khrennikov, 2013; Gonçalves, 2011, 2013).
Choustova (2007a,b), in particular, argued for the introduction of
a quantum-based approach to Financial Theory as a way to incorporate
market cognition dynamics in financial price formation.

The quantum-based approach goes, however, beyond a good match to price
dynamics and turbulence modeling. The growing empirical evidence of
quantum interference signatures in human cognition, when faced with
decision problems, has led to the development of a Quantum Theory-based
Cognitive Science forming a theoretical ground for Econophysics modeling,
with strong implications for Finance (Busemeyer and Franco, 2010;
Busemeyer and Bruza, 2012; Wang and Busemeyer, 2013; Busemeyer and
Wang, 2014; Khrennikov, 2010; Haven and Khrennikov, 2013; Zuo, 2014;
Khrennikov and Basieva, 2014).

The main research problem regarding Quantum Theory-based Cognitive
Science applied to Finance can be expressed as follows: \emph{if}
there is emprical support to the fact that human cognition, in decision
problems, leads to a decision behavior computationally isomorphic
to quantum adaptive computation (Busemeyer and Franco, 2010; Busemeyer
and Bruza, 2012; Wang and Busemeyer, 2013; Busemeyer and Wang, 2014;
Zuo, 2014; Khrennikov and Basieva, 2014; Gonçalves, 2015), \emph{then},
the modeling of financial market dynamics needs to work with models
of behavior that incorporate, in their probabilistic description,
quantum interference terms (Khrennikov, 2010; Haven and Khrennikov,
2013).

This main research problem has led to the growth and development of
research lines on Cognitive Science, working from Quantum Computer
Science and Quantum Information Theory, with direct implications in
Finance and Economics, supporting the expansion of Quantum Econophysics
(Khrennikov, 2010; Haven and Khrennikov, 2013), in particular, in
regards to Finance: opening up the way for research on Quantum Artificial
Intelligence (QuAI) applications to financial market modeling (Gonçalves,
2011, 2013).

The current work contributes to such research by introducing Quantum
Artificial Neural Networks (QuANNs) for financial market dynamics
and volatility risk modeling. In particular, recurrent QuANNs are
used to build a model of financial market dynamics that incorporates
quantum interference and quantum adaptive computation in the probabilistic
description of financial returns. The resulting model shows a quantum-based
selection of adaptive rules with consequences for the market dynamics,
leading to excess kurtosis and turbulence with clustering volatility,
price jumps and statistically significant deviations from Gaussian
distributions, for a wide range of parameters.

The work is divided in two parts that are developed in sections 2
and 3. In section 2, the QuANN model is built, simulated and studied,
while, in section 3, a reflection is provided on the possible role
and contributions of QuAI applied to financial modeling. Regarding
the main work, which is developed in section 2, the structure of this
section is divided in three subsections.

In subsection 2.1, we review a general framework for Classical Econophysics
modeling of financial market price formation in which Farmer's market
making model (Farmer, 2002) is reviewed and combined with multiplicative
components, namely: multiplicative volatility components and a market
polarization component are introduced in the market making model and
linked to trading volume and bullish versus bearish polarization.

In subsection 2.2, we introduce the general formalism of QuANNs, including
main notions that form the groundwork for the financial market model.
In subsection 2.3, we build the financial market model using a Quantum
Neural Automaton (QuNA) structure and simulate the resulting artificial
financial market, addressing its main results in regards to turbulence
and volatility risk, leading to statistically significant deviations
from the Gaussian returns distribution.

In section 3, the problem of deviations from the Gaussian random walk
is addressed in its relation to Econophysics and Nonlinear Stochastic
Models of market dynamics, allowing for a reflection on the possible
contributions of QuAI and QuANNs for establishing a bridge between
the evidence of quantum interference patterns observed in human decision
making and a computational basis for nonlinear probability dynamics
in Finance coming from a linear unitary evolution of networked quantum
computation.

\section{A QuANN-Based Financial Market Model}

\subsection{Price Formation and Financial Returns}

Following Farmer (2002) and Ilinski (2001), financial market price
formation can be linked to unbalanced market orders $M$, where $M>0$
corresponds to an excess demand while $M<0$ to an excess supply,
such that, for a financial risky asset, traded at discrete trading
rounds of duration $\triangle t$, the asset price at $t$, $S(t)$
depends upon the previous price $S(t-\triangle t)$ and the market
orders that arrive during the trading round. A few basic assumptions,
in Classical Econophysics, determine the structure for the relation
between market orders and the new price (Farmer, 2002; Ilinsky, 2001):
\begin{itemize}
\item The price is assumed as a finite increasing function of the previous
price and order size $M(t)$:
\begin{equation}
S(t)=f_{S}\left(S(t-\triangle t),M(t)\right)
\end{equation}

\item If the order size is null $M(t)=0$ the market clears for equal supply
and demand, so that there is no market impact (the price stays unchanged):
\begin{equation}
f_{S}\left(S(t-\triangle t),0\right)=S(t-\triangle t)
\end{equation}

\item There are no arbitrage opportunities associated with a sequence of
trades that sum zero (a repeated trading through a circuit);
\item Gauge invariance with respect to currency units, so that the only
possible combination for prices to enter is $S(t)/S(t-\triangle t)$,
such that:
\begin{equation}
\frac{S(t)}{S(t-\triangle t)}=\frac{f_{S}\left(S(t-\triangle t),M(t)\right)}{S(t-\triangle t)}=F\left(M(t)\right)
\end{equation}

\end{itemize}
The result of these four assumptions is the general form for $F$
in Eq.(3) given by (Farmer, 2002; Ilinsky, 2001):
\begin{equation}
F\left(M(t)\right)=e^{\frac{M(t)}{\lambda}}
\end{equation}
where $\lambda$ is a liquidity parameter, also called market depth
(Farmer, 2002). The result from Eq.(4), replaced in Eq.(3) is the
following dynamical rule:
\begin{equation}
S(t)=S(t-\triangle t)e^{\frac{M(t)}{\lambda}}
\end{equation}
or, taking the logarithms, the log-price rule (Farmer, 2002; Ilinski,
2001):
\begin{equation}
\ln S(t)=\ln S(t-\triangle t)+\frac{M(t)}{\lambda}
\end{equation}
There are two dynamical components to $M(t)$: the sign, which can
either be positive (excess of buy orders) or negative (excess of sell
orders), and the volume of unbalanced market orders, which is linked
to the order size.

Within financial theory, the order size can be worked from a systemic
market dynamics that leads to the formation of consensus clusters
regarding the decision to invest greater or smaller amounts, or, alternatively,
to sell greater or smaller amounts. The adaptive management of exposure
to asset price fluctuation risk, on the part of market agents, given
information that impacts asset value leads to a two-sided aspect of
computation of financial information by the market system: on the
one hand, there is the matter whether each new information is good
(\emph{bullish}) or bad (\emph{bearish}), in terms of asset value,
on the other hand, there is the degree to which new information supports
the decision to buy or sell by different amounts (the market volume
aspect).

A social consensus dynamics coming from market computation can be
linked to consensus clusters affecting the market unbalance, so that
the positive or negative sign can be addressed, within Econophysics,
in terms of a notion of spin. In Physics the spin is a fundamental
degree of freedom of field quanta that behaves like angular momentum,
the spin quantum numbers assume integer and half-integer values, the
most elementary case of half integer spin is the spin-1/2. 

Considering a three dimensional axes system, if a spin-1/2 particle's
spin state is measured along the $z$-axis then there are two fundamental
orientations \emph{spin up} and \emph{spin down}, in Complex Systems
Science these two orientations are assumed and worked mainly from
the statistical mechanics of Ising systems as models of complex systems
(Kauffman, 1993), which constituted early inspiration for Econophysics
models of financial markets (Vaga, 1990; Iiori, 1999; Lux and Marchesi,
1999; Voit, 2001). These models allowed for the study of polarization
in market sentiment, working with the statistical mechanics of Ising
systems, allowing direct connections to Cognitive Science (Voit, 2001).

The market volume, on the other hand, has been addressed, within Financial
Theory, by multiplicative processes (Mandelbrot, et al., 1997; Mandelbrot,
1997), drawing upon Mandelbrot's work on turbulence in Statistical
Mechanics, as reviewed in Mandelbrot (1997). The multiplicative stochastic
processes, worked by Mandelbrot and connected to multifractal geometry,
led to Mandelbrot et al.'s (1997) Multifractal Model of Asset Returns
(MMAR), which also inspired modified versions using multiplicative
stochastic processes with Markov switching in volatility components
(Calvet and Fisher, 2004; Lux, 2008).

Considering Eq.(6), a spin-1/2 like model can be integrated as a binary
component in a multiplicative model that includes market volume, by
way of a multiplictive decomposition of $M(t)$ in a market polarization
component $\sigma(t)=\pm1$, and $N$ trading volume-related volatility
components, so that we obtain:

\begin{equation}
M(t)=\left(\prod_{k=1}^{N}V_{k}(t)\right)\sigma(t)
\end{equation}
where each volatility component $V_{k}(t)$ can assume one of two
values $v_{0}$ or $v_{1}=2-v_{0}$. If $0<v_{0}\leq1$, then $v_{0}$
corresponds to a low volatility state, while $v_{1}$ to a high volatility
state ($v_{0}$ diminishes the returns' value while $v_{1}$ amplifies
the returns value like a lever). The logarithmic returns for the risky
asset, in this approach, are given by:

\begin{equation}
R(t)=\ln\frac{S(t)}{S(t-\triangle t)}=\frac{1}{\lambda}\left(\prod_{k=1}^{N}V_{k}(t)\right)\sigma(t)
\end{equation}

The binary structure assumed for the $N$ components plus the market
polarization, makes this model a good starting point for QuANN applications,
since QuANNs also work from a binary computational basis to address
neural firing patterns. 

On the other hand, QuANNs open up the possibility for dealing with
the multiplicative models in such a way that the probabilities, rather
than being introduced from a top-down ex-ante fixed state-transition
probability distribution, change from trading round to trading round,
being the result of the quantum computational process introduced for
each returns' component.

QuANNs also allow one to incorporate the empirical evidence that human
cognition, when addressing decision between alternatives, follows
a dynamics that is computationally isomorphic to quantum computation
applied to Decision Science, leading to interference effects with
an expression in decision frequencies (probabilities), which means
that, when considering probabilities for human behavior, the theoretical
framework of networked quantum computation may be more appropriate
for the dynamical modeling of human systems.

In the quantum description, Eq.(8) will be expressed in operator form
on an appropriate Hilbert space, with the returns operator eigenvalues
being addressed from the QuANN structure, which works with \emph{quantum
bits} (\emph{qubits}), whose computational basis states decribe the
neuron's firing pattern in terms of firing (ON) and non-firing (OFF)\footnote{This degree of freedom behaves like spin, so that the neuron's associated
\emph{qubit} can also be approached in terms of a spin-1/2 model.}. In order to build the market model, however, we need to introduce,
first, a general framework for QuANNs which will then be applied to
the risky asset price dynamics modeling.

\subsection{Quantum Artificial Neural Networks}

The connection between Quantum Computer Science and ANNs has been
object of research since the 1990s, in particular, in what regards
quantum associative memory, quantum parallel processing, extension
of classical ANN schemes, as well as computational complexity and
efficiency of QuANNs over classical ANNs (Chrisley, 1995; Kak, 1995;
Menneer and Narayanan, 1995; Behrman \emph{et al.}, 1996; Menneer,
1998; Ivancevic and Ivancevic, 2010; Gonçalves, 2015).

Mathematically, a classical ANN with a binary firing pattern can be
defined as an artificial networked computing system comprised of a
directed graph with the following additional structure (McCulloch
and Pitts, 1943; Müller \emph{et al.}, 1995):
\begin{itemize}
\item A binary alphabet $\mathbb{A}_{2}=\left\{ 0,1\right\} $ associated
to each neuron describing the neural activity, with 0 corresponding
to a non-firing neural state and 1 to a firing neural state, so that
the firing patterns of a neural network with $N$ neurons are expressed
by the set of all binary strings of length $N$: $\mathbb{A}_{2}^{N}=\left\{ s_{1}s_{2}...s_{N}:s_{k}\in\mathbb{A}_{2},k=1,2,...,N\right\} $;
\item A real-valued weight associated with each neural link, expressing
the strength and type of neural connection;
\item A transfer function which determines the state transition of the neuron
and that depends upon the state of its incident neurons, the weight
associated with each incoming neural links and an activation threshold
that can be specific for each neuron.
\end{itemize}
A quantum version of ANNs, on the other hand, can be defined as a
directed graph with a networked quantum computing structure, such
that (Gonçalves, 2015):
\begin{itemize}
\item To each neuron is associated a two-dimensional Hilbert Space $\mathcal{H}_{2}$
spanned by the computational basis $\mathcal{B}_{2}=\left\{ \left|0\right\rangle ,\left|1\right\rangle \right\} $,
where $\left|0\right\rangle ,\left|1\right\rangle $ are ket vectors
(in Dirac's \emph{bra-ket} notation for Quantum Mechanics' vector-based
formalism using Hilbert spaces\footnote{We use the vector representation convention introduced by Dirac (1967)
for Hilbert spaces, assumed and used extensively in Quantum Mechanics.
In this case, a \emph{ket vector}, represented as $\left|a\right\rangle $,
is a column vector of complex numbers while a \emph{bra vector}, represented
as $\left\langle a\right|$, is the conjugate transpose of $\left|a\right\rangle $,
that is: $\left\langle a\right|=\left|a\right\rangle ^{\dagger}$.
The Hilbert space inner product is represented as $(\left|a\right\rangle ,\left|b\right\rangle )=\left\langle a|b\right\rangle $.
The outer product is, in turn, given by $\left|a\right\rangle \left\langle b\right|$.
A projection operator corresponds to an operator of the form $\hat{P}_{a}=\left|a\right\rangle \left\langle a\right|$
which acts on any ket $\left|b\right\rangle $ as $\hat{P}_{a}\left|b\right\rangle =\left\langle a|b\right\rangle \left|a\right\rangle $.}), where $\left|0\right\rangle $ encodes a non-firing neural dynamics
and $\left|1\right\rangle $ encodes a firing neural dynamics;
\item To a neural network, comprised of $N$ neurons, is associated the
tensor product of $N$ copies of $\mathcal{H}_{2}$, so that the neural
network's Hilbert space is the space $\mathcal{H}_{2}^{\otimes N}$
spanned by the basis $\mathcal{B}_{2}^{\otimes N}=\left\{ \left|\mathbf{s}\right\rangle :\mathbf{s}\in\mathbb{A}_{2}^{N}\right\} $
which encodes all the alternative firing patterns of the neurons;
\item The general neural configuration state of the neural network is characterized
by a normalized ket vector $\left|\psi\right\rangle \in\mathcal{H}_{2}^{\otimes N}$
expanded in the neural firing pattterns' basis $\mathcal{B}_{2}^{\otimes N}$
as:
\begin{equation}
\left|\psi\right\rangle =\sum_{\mathbf{s}\in\mathbb{A}_{2}^{N}}\psi(\mathbf{s})\left|\mathbf{s}\right\rangle 
\end{equation}
with the normalization condition:
\begin{equation}
\sum_{\mathbf{s}\in\mathbb{A}_{2}^{N}}|\psi(\mathbf{s})|^{2}=1
\end{equation}

\item The neural network has an associated neural links state transition
operator $\hat{L}_{Net}$ such that, given an input neural state $\left|\psi_{in}\right\rangle $,
the operator transforms the input state for the neural network in
an output state $\left|\psi_{out}\right\rangle $, reflecting, in
this operation, the neural links for the neural network, so that each
neuron has an associated structure of unitary operators that is conditional
on its input neurons:
\begin{equation}
\left|\psi_{out}\right\rangle =\hat{L}_{Net}\left|\psi_{in}\right\rangle 
\end{equation}

\end{itemize}
The output state of a QuANN shows, in general, complex quantum correlations
so that the quantum dynamics of a single neuron may depend in a complex
way on the entire neural network's configuration (Gonçalves, 2015).
Considering the neurons $n_{1},...,n_{N}$ for a $N$-neuron neural
network, the $\hat{L}_{Net}$ operator can be expressed as a product
of each neuron's neural links operator following the ordered sequence
$n_{1},...,n_{N}$, where neuron $n_{1}$ is the first to be updated
and $n_{N}$ the last (that is, following the activation sequence\footnote{For some QuANNs it is possible to consider the action of the operators
conjointly and to introduce, in one single neural links operator,
a transformation of multiple neurons' states, taking advantage of
parallel quantum computation (Gonçalves, 2015).}):
\begin{equation}
\hat{L}_{Net}=\hat{L}_{N}...\hat{L}_{2}\hat{L}_{1}
\end{equation}
each neuron's neural links operator is a quantum version of an activation
function, with the following structure for the $k$-th neuron: 
\begin{equation}
\hat{L}_{k}=\sum_{\mathbf{s}\in\mathbb{A}_{2}^{k-1},\mathbf{s'}\in\mathbb{A}_{2}^{N-k}}\left|\mathbf{s}\right\rangle \left\langle \mathbf{s}\right|\otimes L_{k}(\mathbf{s}_{in})\otimes\left|\mathbf{s'}\right\rangle \left\langle \mathbf{s'}\right|
\end{equation}
where $\mathbf{s}_{in}$ is a substring, taken from the binary word
$\mathbf{s}\mathbf{s'}$, that matches in $\mathbf{s}\mathbf{s'}$
the activation pattern for the input neurons of $n_{k}$, under the
neural network's architecture, in the same order and binary sequence
as it appears in $\mathbf{s}\mathbf{s'}$, $L_{k}(\mathbf{s}_{in})$
is a neural links function that maps the input substring to a unitary
operator on the two-dimensional Hilbert space $\mathcal{H}_{2}$,
this means that, for different configurations of the neural network,
the neural links operator for the $k$-th neuron $\hat{L}_{k}$ assigns
a corresponding unitary operator that depends upon the activation
pattern of the input neurons.

The neural links operators incorporate the local structure of neural
connections so that there is a unitary state transition for the neuron
(a quantum computation) conditional upon the firing pattern of its
input neurons.

Now, an arbitrary unitary operator on a single-\emph{qubit} Hilbert
space $\mathcal{H}_{2}$ is a member of the unitary group U(2) and
can be derived from a specific Hamiltonian operator structure (Greiner
and Müller, 2001), so that we have, for a QuANN, a conditional unitary
state transition:

\begin{equation}
L_{k}(\mathbf{s}_{in}):=e^{-\frac{i}{\hbar}\triangle t\hat{H}_{\mathbf{s}_{in}}}
\end{equation}
where the neuron's associated Hamiltonian operator $\hat{H}_{\mathbf{s}_{in}}$
is conditional on the input neurons' firing pattern $\mathbf{s}_{in}$
and given by the general structure: 
\begin{equation}
\hat{H}_{\mathbf{s}_{in}}=-\frac{\omega(\mathbf{s}_{in})}{2}\hbar\hat{1}+\theta(\mathbf{s}_{in})\sum_{j=1}^{3}u_{j}(\mathbf{s}_{in})\frac{\hbar}{2}\hat{\sigma}_{j}
\end{equation}
where $\hbar$ is the reduced Planck constant\footnote{$1.054571800(13)\times10^{-34}$Js},
$\theta(\mathbf{s}_{in})$, $\omega(\mathbf{s}_{in})$ are measured
in radians per second and depend upon the neural configuration for
the input neurons, $\hat{1}$ is the unit operator on $\mathcal{H}_{2}$,
the $u_{j}(\mathbf{s}_{in})$ terms are the components of a real unit
vector $\mathbf{u}(\mathbf{s}_{in})$ and $\hat{\sigma}_{j}$ are
Pauli's operators\footnote{The terms $(\hbar/2)\hat{\sigma}_{j}$, in the Hamiltonian, are equivalent
to the spin operators for a spin-1/2 system (Leggett, 2002).}: 
\begin{equation}
\hat{\sigma}_{1}=\left|0\right\rangle \left\langle 1\right|+\left|1\right\rangle \left\langle 0\right|=\left(\begin{array}{cc}
0 & 1\\
1 & 0
\end{array}\right)
\end{equation}
\begin{equation}
\hat{\sigma}_{2}=-i\left|0\right\rangle \left\langle 1\right|+i\left|1\right\rangle \left\langle 0\right|=\left(\begin{array}{cc}
0 & -i\\
i & 0
\end{array}\right)
\end{equation}
\begin{equation}
\hat{\sigma}_{3}=\left|0\right\rangle \left\langle 0\right|-\left|1\right\rangle \left\langle 1\right|=\left(\begin{array}{cc}
1 & 0\\
0 & -1
\end{array}\right)
\end{equation}
Replacing Eq.(15) in Eq.(14) and expanding we obtain: 
\begin{equation}
\begin{aligned}L_{k}(\mathbf{s}_{in})=e^{-\frac{i}{\hbar}\triangle t\hat{H}_{\mathbf{s}_{in}}}=\\
=e^{i\frac{\omega(\mathbf{s}_{in})\triangle t}{2}}\left[\cos\left(\frac{\theta(\mathbf{s}_{in})\triangle t}{2}\right)\hat{1}-i\sin\left(\frac{\theta(\mathbf{s}_{in})\triangle t}{2}\right)\sum_{j=1}^{3}u_{j}(\mathbf{s}_{in})\hat{\sigma}_{j}\right]
\end{aligned}
\end{equation}
where $\hat{1}=\left|0\right\rangle \left\langle 0\right|+\left|1\right\rangle \left\langle 1\right|$
is the unit operator on $\mathcal{H}_{2}$. The operator in Eq.(19)
is comprised of the product of a phase transformation $\exp\left(i\omega(\mathbf{s}_{in})\triangle t/2\right)$
and a rotation operator defined as (Greiner and Müller, 2001; Nielsen
and Chuang, 2003):
\begin{equation}
\hat{R}_{\mathbf{u}(\mathbf{s}_{in})}\left[\theta(\mathbf{s}_{in}),\triangle t\right]=\cos\left(\frac{\theta(\mathbf{s}_{in})\triangle t}{2}\right)\hat{1}-i\sin\left(\frac{\theta(\mathbf{s}_{in})\triangle t}{2}\right)\sum_{j=1}^{3}u_{j}(\mathbf{s}_{in})\hat{\sigma}_{j}
\end{equation}
An arbitrary single-\emph{qubit} unitary operator (a quantum logic
gate on a \emph{qubit}) can, thus, be expressed by the product (Nielsen
and Chuang, 2003):
\begin{equation}
e^{-\frac{i}{\hbar}\hat{H}_{\mathbf{s}_{in}}\triangle t}=\exp\left(i\frac{\omega(\mathbf{s}_{in})\triangle t}{2}\right)\hat{R}_{\mathbf{u}(\mathbf{s}_{in})}\left[\theta(\mathbf{s}_{in})\triangle t\right]
\end{equation}
This means that the transfer function of classical ANNs is replaced,
for QuANNs, by phase transforms and rotations of the neuron's quantum
state conditional upon the firing pattern of the input neurons\footnote{This leads to quantum correlations that reflect the neural network's
structure (Gonçalves, 2015).}.

Now, given an operator $\hat{O}$ on the neural network's Hilbert
space $\mathcal{H}_{2}^{\otimes N}$ expanded as: 
\begin{equation}
\hat{O}=\sum_{\mathbf{s},\mathbf{s'}\in\mathbb{A}_{2}^{N}}O_{\mathbf{s},\mathbf{s'}}\left|\mathbf{s}\right\rangle \left\langle \mathbf{s'}\right|
\end{equation}
taking the inner product between a normalized ket vector $\left|\psi\right\rangle $
and the transformed vector $\hat{O}\left|\psi\right\rangle $ yields
:
\begin{equation}
\begin{aligned}\left(\left|\psi\right\rangle ,\hat{O}\left|\psi\right\rangle \right)=\left\langle \psi\left|\hat{O}\right|\psi\right\rangle =\\
=\sum_{\mathbf{s},\mathbf{s'}\in\mathbb{A}_{2}^{N}}O_{\mathbf{s},\mathbf{s'}}\left\langle \mathbf{s'}|\psi\right\rangle \left\langle \psi|\mathbf{s}\right\rangle =\\
=\sum_{\mathbf{s},\mathbf{s'}\in\mathbb{A}_{2}^{N}}O_{\mathbf{s},\mathbf{s'}}\psi(\mathbf{s'})\psi(\mathbf{s})^{*}
\end{aligned}
\end{equation}
For Hermitian operators obeying the relation:
\begin{equation}
O_{\mathbf{s},\mathbf{s'}}\left\langle \mathbf{s'}|\psi\right\rangle \left\langle \psi|\mathbf{s}\right\rangle \delta_{\mathbf{s},\mathbf{s'}}
\end{equation}
given that the state vector is normalized, if this relation is verified,
then Eq.(23) yields a classical expectation in which the amplitudes
in square modulus $|\psi(\mathbf{s})|^{2}$ are equivalent to decision
weights associated with each alternative value on the diagonal of
the operator's matrix representation:
\begin{equation}
\left\langle \hat{O}\right\rangle _{\psi}=\left\langle \psi\left|\hat{O}\right|\psi\right\rangle =\sum_{\mathbf{s}\in\mathbb{A}_{2}^{N}}O_{\mathbf{s},\mathbf{s}}|\psi(\mathbf{s})|^{2}
\end{equation}
so that, for a neural network in the state $\left|\psi\right\rangle $,
the neural activity can be described by the value $O_{\mathbf{s},\mathbf{s}}$
with an associated weight of $|\psi(\mathbf{s})|^{2}$.

In the case of Econophysics, as well as Game Theory applications,
one usually assumes that the social system tends to the alternatives
in proportion to the corresponding decision weights, such that one
can associate a probability measure for the system to follow each
alternative as numerically coincident to the corresponding decision
weight. This is akin to Game Theory's notion of mixed strategy, in
the sense that each player can be characterized by a fixed mixed strategy
and play probabilistically according to the mixed strategy's weights.

While the probability of a player's behavior is zero or one after
play, the decision weights remain the same, in the case of Game Theory
this means that the Nash equilibrium does not change, being available
as a cognitive strategic scheme for further plays (Nash, 1951). In
applications of QuANNs to social systems this means that one needs
to work with either an Everettian interpretation of Quantum Theory,
or with a Bohmian interpretation\footnote{Since these are the two lines of interpretation that do not assume
a state vector collapse.}.

The Bohmian interpretation is often assumed by researchers dealing
with Econophysics (Choustova, 2007a,b; Khrennikov, 2010; Haven and
Khrennikov, 2013), in particular, when one wishes to address the amplitudes
in square modulus $|\psi(\mathbf{s})|^{2}$ in terms of economic forces
linked to emergent degrees of freedom that tend to make the system
follow certain paths probabilistically (a quantum-based probabilistic
version of Haken's slaving principle applied to economic and financial
systems (Haken, 1977)).

The Everettian line of interpretations has, since its initial proposal
by Everett (1957, 1973), been directly linked to a Cybernetics' paradigmatic
basis incorporating both Automata Theory and Information Theory (Gonçalves,
2015), a point that comes directly from Everett's original work on
Quantum Mechanics, that is further deepened by Deutsch's work on Quantum
Computation (Deutsch, 1985), and, later, on Quantum Decision Theory
(Deutsch, 1999; Wallace, 2002, 2007).

There are actually different perspectives from different authors on
Everett's original proposal (Bruce, 2004). Formally, the proposal
is close to Bohm's, including the importance attributed to computation
and to Information Theory, however, systemically, Bohm and Everett
are very distinct in the hypotheses they raise: for Bohm the state
vector is assumed to represent a statistical average of an underlying
information field's sub-quantum dynamics (Bohm, 1984; Bohm and Hiley,
1993), Everett (1957, 1973) assumes the geometry of the Hilbert space
as the correct description of the fundamental dynamics of fields and
systems.

Considering QuANNs, under Everett's approach, and working from the
state vector, we can introduce the set of projection operators onto
the basis $\mathcal{B}_{2}^{\otimes N}$, $\mathcal{P}=\left\{ \hat{P}_{\mathbf{s}}=\left|\mathbf{s}\right\rangle \left\langle \mathbf{s}\right|:\mathbf{s}\in\mathbb{A}_{2}^{N}\right\} $
where each operator has the matrix representation $P_{\mathbf{s},\mathbf{s'}}=\delta_{\mathbf{s},\mathbf{s'}}$,
these operators form a complete set of orthogonal projectors, since
their sum equals the unit operator on the Hilbert space $\mathcal{H}_{2}^{\otimes N}$,
$\sum_{\mathbf{s}\in\mathbb{A}_{2}^{N}}\hat{P}_{\mathbf{s}}=\hat{1}^{\otimes N}$,
and they are mutually exclusive, that is, the product of two of these
operators obeys the relation $\delta_{\mathbf{s},\mathbf{s'}}\hat{P}_{\mathbf{s}}\hat{P}_{\mathbf{s'}}$.

A projection operator can represent a projective computation, by the
neural network, of an alternative neural firing pattern for the network.
The general state vector in Eq.(9) can, thus, be expressed as a sum
of projections, that is, the neural network's quantum state has a
projective expression over each alternative neural configuration simultaneously,
corresponding to a simultaneous systemic projective activity over
all alternatives:
\begin{equation}
\left|\psi\right\rangle =\sum_{\mathbf{s}\in\mathbb{A}_{2}^{N}}\hat{P}_{\mathbf{s}}\left|\psi\right\rangle 
\end{equation}
Each alternative neural configuration corresponds to an orthogonal
dimension of the $2^{N}$ dimensional Hilbert space $\mathcal{H}_{2}^{\otimes N}$,
a dimension that is spanned by a corresponding basis vector in $\mathcal{B}_{2}^{\otimes N}$,
which means that the quantum system (in our case, the QuANN) projects
simultaneously over each (orthogonal) dimension of systemic activity
(corresponding, in our case, to each alternative neural pattern) weighing
each dimension. The wheight of the projection over a given dimension
(a given pattern of systemic activity) in the system's state can be
worked from a notion of norm. Using the Hilbert space's inner product
structure, we can work with the squared norm of the projected vector,
which leads to:
\begin{equation}
\left\Vert \hat{P}_{\mathbf{s}}\left|\psi\right\rangle \right\Vert ^{2}=(\hat{P}_{\mathbf{s}}\left|\psi\right\rangle ,\hat{P}_{\mathbf{s}}\left|\psi\right\rangle )=\left\langle \psi\left|\hat{P}_{\mathbf{s}}^{\dagger}\hat{P}_{\mathbf{s}}\right|\psi\right\rangle =|\psi(\mathbf{s})|^{2}
\end{equation}
Systemically, this last equation can be interpreted as expressing
that the weight of the projection $\hat{P}_{\mathbf{s}}$, in the
system's projective dynamics, is equal to $|\psi(\mathbf{s})|^{2}$.
In this sense, each orthogonal dimension corresponds to a distinct
pattern of activity that is projectively computed by the system.

On the other hand, for a large ensemble of QuANNs with the same structure
and in the same state, the statistical weight associated to the projection
operator $\hat{P}_{\mathbf{s}}$, expressed by the ensemble average
$\left\langle \hat{P}_{\mathbf{s}}\right\rangle $, coincides with
the projection weight $|\psi(\mathbf{s})|^{2}$ associated to the
neural state projection $\hat{P}_{\mathbf{s}}\left|\psi\right\rangle $,
thus, the statistical interpretation comes directly from the projective
structure of the state vector. Indeed, let us consider a statistical
ensemble of $M$ QuANNs such that each QuANN has the same number of
neurons $N$ and the same architecture, let us, further, assume that
each neural network is characterized by some quantum neural state
$\left|\psi_{k}\right\rangle $, with $k=1,2,...,M$, the ensemble
state can be represented by a statistical density operator:
\begin{equation}
\hat{\rho}=\frac{1}{M}\sum_{k=1}^{M}\left|\psi_{k}\right\rangle \left\langle \psi_{k}\right|
\end{equation}
the statistical average of an operator $\hat{O}$ on the Hilbert space
$\mathcal{H}_{2}^{\otimes N}$ is given by (Bransden and Joachain,
2000):

\begin{equation}
\begin{aligned}\left\langle \hat{O}\right\rangle _{\hat{\rho}}=Tr(\hat{O}\hat{\rho})=\\
=\frac{1}{M}\sum_{k=1}^{M}\sum_{\mathbf{s},\mathbf{s'}\in\mathbb{A}_{2}^{N}}O_{\mathbf{s},\mathbf{s'}}\left\langle \mathbf{s'}|\psi_{k}\right\rangle \left\langle \psi_{k}|\mathbf{s}\right\rangle \\
=\frac{1}{M}\sum_{k=1}^{M}\left\langle \psi_{k}\left|\hat{O}\right|\psi_{k}\right\rangle =\frac{1}{M}\sum_{k=1}^{M}\left\langle \hat{O}\right\rangle _{\psi_{k}}
\end{aligned}
\end{equation}
for a projector on the neural basis we get the ensemble average: 
\begin{equation}
\left\langle \hat{P}_{\mathbf{s}}\right\rangle _{\hat{\rho}}=\frac{1}{M}\sum_{k=1}^{M}\left\langle \psi_{k}\left|\hat{P}_{\mathbf{s}}\right|\psi_{k}\right\rangle =\frac{1}{M}\sum_{k=1}^{M}\left|\psi_{k}(\mathbf{s})\right|^{2}
\end{equation}
Now, if all the members of the ensemble are in the same neural state
$\left|\psi_{k}\right\rangle =\left|\psi\right\rangle $ for each
$k=1,...,M$ the whole statistical weight that is placed on the projection
coincides exactly with $|\psi(\mathbf{s})|^{2}$ so that the ensemble
average of the projection coincides numerically with the degree to
which the system projects over the dimension corresponding to the
neural pattern $\left|\mathbf{s}\right\rangle $ (the projetion norm),
that is, there is a numerical coincidence between $\left\Vert \hat{P}_{\mathbf{s}}\left|\psi\right\rangle \right\Vert ^{2}$
and $\left\langle \hat{P}_{\mathbf{s}}\right\rangle _{\hat{\rho}}$:
\begin{equation}
\left\langle \hat{P}_{\mathbf{s}}\right\rangle _{\hat{\rho}}=\frac{1}{M}\sum_{k=1}^{M}\left|\psi(\mathbf{s})\right|^{2}=\left|\psi(\mathbf{s})\right|^{2}
\end{equation}

Thus, an ensemble of QuANNs with the same structure, characterized
by the same quantum state $\left|\psi\right\rangle $, has a statistical
weight for each projection coincident with the norm of the projection,
so that this norm has a statistical expression once we consider an
ensemble of systems with the same structure and characterized by the
same state. This is similar to the argument that is made around repeated
independent\footnote{In the case of QuANNs this pressuposes the non-interaction between
the ensemble elements, appealing to a description of a statistical
random sample.} and identically prepared experiments leading to a statistical distribution
that shows the markers of the underlying quantum dynamics, in that
case, we also see a statistical ensemble marker (considering an ensemble
of experiments with the same state vector) that recovers the projection
norm structure in the statistical distribution.

The experiments, in the case of human systems, have led to the finding
of the same computational properties and projective dynamics present
in the quantum systems (Busemeyer and Franco, 2010; Busemeyer and
Bruza, 2012; Wang and Busemeyer, 2013; Busemeyer and Wang, 2014),
a finding that comes from the statistical distribution of the experiments.

In an Econophysics setting, the projective dynamics can be addressed
as a cognitive projection such that the projection norm corresponds
to the decision weight placed on that alternative\footnote{In the quantum computational setting, under the Everettian line, the
projective structure for QuANNs can be considered as a computational
projection such that each Hilbert space dimension, corresponding to
a different neural pattern, is computed simultaneously with an associated
weight (given by the norm of the projection), having a computational
expression in the system's quantum processing and a statistical correspondence
in the neural activity pattern of an ensemble of QuANNs with the same
structure and in the same state (assuming non-interaction between
different ensemble elements). 

In the case of physical systems, the projective dynamics, interpreted
computationally, leads to a physical expression of the system at multiple
dimensions of systemic activity, a point which was interpreted by
DeWitt (1970) under the notion of Many Worlds of a same Universe,
where each World corresponds to an entire configuration of the Universe
matching a corresponding orthogonal dimension of an appropriate Hilbert
space where observers and systems are correlated (entanglement).

In the case of applications to human decision-making, the orthogonal
dimensions can be assumed to correspond to alternative decision scenarios
evaluated by the decision-maker and supporting his/her choice.}. The QuANN state transition has an implication in the projection
weights, in the sense that given the state transition:

\begin{equation}
\left|\psi_{out}\right\rangle =\hat{L}_{Net}\left|\psi_{in}\right\rangle =\sum_{\mathbf{s}\in\mathbb{A}_{2}^{N}}\psi_{out}(\mathbf{s})\left|\mathbf{s}\right\rangle ,
\end{equation}
the output amplitudes are given by:
\begin{equation}
\psi_{out}(\mathbf{s})=\sum_{\mathbf{s'}\in\mathbb{A}_{2}^{N}}\left\langle \mathbf{s}\left|\hat{L}_{Net}\right|\mathbf{s'}\right\rangle \left\langle \mathbf{s'}|\psi_{in}\right\rangle =\sum_{\mathbf{s}'\in\mathbb{A}_{2}^{N}}L_{Net}(\mathbf{s},\mathbf{s'})\psi_{in}(\mathbf{s}')
\end{equation}
with $L_{Net}(\mathbf{s},\mathbf{s'})=\left\langle \mathbf{s}\left|\hat{L}_{Net}\right|\mathbf{s'}\right\rangle $.
Eq.(33) means that the following change in the projections' norms
takes place:
\begin{equation}
\begin{aligned}\left\Vert \hat{P}_{\mathbf{s}}\left|\psi_{in}\right\rangle \right\Vert ^{2}=|\psi_{in}(\mathbf{s})|^{2}\rightarrow\\
\rightarrow\left\Vert \hat{P}_{\mathbf{s}}\left|\psi_{out}\right\rangle \right\Vert ^{2}=|\psi_{out}(\mathbf{s})|^{2}=\left|\sum_{\mathbf{s}'\in\mathbb{A}_{2}^{N}}L_{Net}(\mathbf{s},\mathbf{s'})\psi_{in}(\mathbf{s}')\right|^{2}
\end{aligned}
\end{equation}
The sum within the square modulus is a source of quantum interference
at the projection norm level.

An iterative scheme with the repeated application of the neural network
operator $\hat{L}_{Net}$ leads to a sequence of quantum neural states
$\left|\psi(t)\right\rangle $. Expanding the complex numbers associated
to the quantum amplitudes: 
\begin{equation}
\left\langle \mathbf{s}|\psi(t)\right\rangle =\psi(\mathbf{s},t)=\sqrt{A(\mathbf{s},t)}+i\sqrt{B(\mathbf{s},t)}
\end{equation}
we can express the dynamical variables $A(\mathbf{s},t)$ and $B(\mathbf{s},t)$
in terms of a dynamical nonlinear state transition rule: 
\begin{equation}
A(\mathbf{s},t)=\left[\textrm{Re}\left(\sum_{\mathbf{s}'\in\mathbb{A}_{2}^{N}}L_{Net}(\mathbf{s},\mathbf{s'})\psi(\mathbf{s}',t-\triangle t)\right)\right]^{2}
\end{equation}
\begin{equation}
B(\mathbf{s},t)=\left[\textrm{Im}\left(\sum_{\mathbf{s}'\in\mathbb{A}_{2}^{N}}L_{Net}(\mathbf{s},\mathbf{s'})\psi(\mathbf{s}',t-\triangle t)\right)\right]^{2}
\end{equation}
which leads to a $2^{N+1}$ system of nonlinear equations, from where
it follows that the probability associated to a given neural firing
configuration, worked from the expected projection (in accordance
with the ensemble average), is given by the sum of the two dynamical
variables:
\begin{equation}
Prob[\mathbf{s},t]=A(\mathbf{s},t)+B(\mathbf{s},t)
\end{equation}

This establishes a bridge between Nonlinear Dynamical Systems Theory
and quantum processing by QuANNs, with implications for financial
modeling. Indeed, while, traditionally, in financial econometrics
one can see the distinction between a stochastic process (be it linear
or nonlinear) and a deterministic nonlinear dynamical system, in the
case of QuANNs applied to financial modeling they synthesize both
approaches (stochastic and deterministic nonlinear), since the quantum
state transition equations have a corresponding expression in a nonlinear
deterministic dynamical system for probability measures assigned to
the QuANN's statistical description via the correspondence between
the projection norm dynamics and the statistical expectation associated
to the projection operator.

The QuANNs application to financial modeling, thus, allows us to address
the problem of simulating the resulting system dynamics that comes
from a human cognition where interference patterns are found in the
probabilistic description of human behavior.

\subsection{A Quantum Market Model}

Considering the financial case, a quantum regime switching model for
the $N$ volatility components plus the market polarization component,
introduced in subsection 2.1, can be addressed through a Quantum Neural
Automaton (QNA), defined as a one dimensional lattice with a QuANN
associated to each lattice site, in this case we assume the lattice
to have $N+1$ sites and to each site $k$, for $k=1,2,...,N+1$,
is associated a QuANN with the following structure:
\begin{equation}
\begin{array}{ccc}
n_{1}(k) & \rightarrow & n_{2}(k)\\
 & \nwarrow & \downarrow\uparrow\\
 &  & n_{3}(k)
\end{array}
\end{equation}
The corresponding Hilbert space for each such neural network $\mathcal{H}_{Net}(k)$
is $\mathcal{H}_{2}^{\otimes3}$ that is $\mathcal{H}_{Net}(k)=\mathcal{H}_{2}^{\otimes3}$,
for $k=1,2,...,N+1$, with the general basis vector $\left|s_{1}s_{2}s_{3}\right\rangle $,
such that $s_{1}$ characterizes the activity pattern of the first
neuron ($n_{1}(k)$), $s_{2}$ characterizes the second neuron ($n_{2}(k)$)
and $s_{3}$ characterizes the activity pattern of the third neuron
($n_{3}(k)$). In what follows, the neuron $n_{3}(k)$ encodes the
market state for the corresponding component, $n_{1}(k)$ encodes
the new market conditions supporting the corresponding component's
dynamics and $n_{2}(k)$ addresses the computation of the synchronization
pattern between $n_{3}(k)$ (the market state for the component) and
$n_{1}(k)$ (the new market conditions). The QNA Hilbert space $\mathcal{H}_{QNA}=\bigotimes_{k=1}^{N+1}\mathcal{H}_{Net}(k)$
is the tensor product of $N+1$ copies of the Hilbert space $\mathcal{H}_{2}^{\otimes3}$.
Assuming this structure for the QNA, we now begin by addressing the
local neural dynamics and its financial interpretation.

\subsubsection{Local Neural Dynamics}

Since the third neuron firing patterns encode the market state of
the corresponding component, for the $N$ volatility components, we
have the neural network market volatility operator on $\mathcal{H}_{2}^{\otimes3}$:
\begin{equation}
\hat{O}_{V}\left|s_{1}s_{2}0\right\rangle =v_{0}\left|s_{1}s_{2}0\right\rangle 
\end{equation}
\begin{equation}
\hat{O}_{V}\left|s_{1}s_{2}1\right\rangle =v_{1}\left|s_{1}s_{2}1\right\rangle 
\end{equation}
while for the market polarization component we have the neural network
market polarization operator:
\begin{equation}
\hat{O}_{P}\left|s_{1}s_{2}0\right\rangle =-1\left|s_{1}s_{2}0\right\rangle 
\end{equation}
\begin{equation}
\hat{O}_{P}\left|s_{1}s_{2}1\right\rangle =+1\left|s_{1}s_{2}1\right\rangle 
\end{equation}
Since, as defined previously, $v_{0}\leq v_{1}$, for a volatility
neural network, when the third neuron fires we have a high volatility
state, and when it does not fire we have a low volatility state. For
the market polarization neural network, when the third neuron fires
we have a \emph{bullish} market state and when it does not fire we
have a \emph{bearish} market state.

Eqs.(40) to (43) show that both operators depend only on the third
neuron's firing pattern, which means that, using Dirac's \emph{bra-ket}
notation, they can be expanded, respectively, as:

\begin{equation}
\hat{O}_{V}=v_{0}\left(\sum_{s_{1},s_{2}\in\mathbb{A}_{2}}\left|s_{1}s_{2}0\right\rangle \left\langle s_{1}s_{2}0\right|\right)+v_{1}\left(\sum_{s_{1},s_{2}\in\mathbb{A}_{2}}\left|s_{1}s_{2}1\right\rangle \left\langle s_{1}s_{2}1\right|\right)
\end{equation}
\begin{equation}
\hat{O}_{P}=-\left(\sum_{s_{1},s_{2}\in\mathbb{A}_{2}}\left|s_{1}s_{2}0\right\rangle \left\langle s_{1}s_{2}0\right|\right)+\left(\sum_{s_{1},s_{2}\in\mathbb{A}_{2}}\left|s_{1}s_{2}1\right\rangle \left\langle s_{1}s_{2}1\right|\right)
\end{equation}

Now, the neural network follows a closed loop starting at the market
state neuron ($n_{3}(k)$) and ending at the market state neuron.
The final state transition amplitudes and the underlying financial
dynamics will depend upon the intermediate transformations which may
change the profile of the corresponding component's state transition
structure.

To address the neural dynamics and its relation with the financial
market dynamics we need to introduce the neural links operators and
follow the loop, starting at $n_{3}(k)$ and ending at $n_{3}(k)$.
Considering, then, the first neural link $n_{3}(k)\rightarrow n_{1}(k)$,
we introduce the following neural network operator for the neuron
$n_{1}(k)$:
\begin{equation}
\hat{L}_{1}=\sum_{s\in\mathbb{A}_{2}}e^{-\frac{i}{\hbar}\triangle t\hat{H}_{0}}\otimes\left|s\right\rangle \left\langle s\right|\otimes\left|0\right\rangle \left\langle 0\right|+\sum_{s'\in\mathbb{A}_{2}}e^{-\frac{i}{\hbar}\triangle t\hat{H}_{1}}\otimes\left|s'\right\rangle \left\langle s'\right|\otimes\left|1\right\rangle \left\langle 1\right|
\end{equation}
using Eq.(19) we need to define the angles $\theta(0)$, $\theta(1)$,
$\omega(0)$, $\omega(1)$ and the vectors $\mathbf{u}(0)$, $\mathbf{u}(1)$,
we set, in this case:
\begin{equation}
\frac{\theta(0)\triangle t}{2}=\phi+\frac{\pi}{2},\:\frac{\theta(1)\triangle t}{2}=\phi
\end{equation}
\begin{equation}
\frac{\omega(0)\triangle t}{2}=\pi,\:\frac{\omega(1)\triangle t}{2}=\frac{\pi}{2}
\end{equation}
\begin{equation}
\mathbf{u}(0)=\mathbf{u}(1)=(1,0,0)
\end{equation}
leading to the following operator structure:
\begin{equation}
e^{-\frac{i}{\hbar}\triangle t\hat{H}_{0}}=\sin\left(\phi\right)\hat{1}+i\cos\left(\phi\right)\hat{\sigma}_{1}=\left(\begin{array}{cc}
\sin\left(\phi\right) & i\cos\left(\phi\right)\\
i\cos\left(\phi\right) & \sin\left(\phi\right)
\end{array}\right)
\end{equation}
\begin{equation}
e^{-\frac{i}{\hbar}\triangle t\hat{H}_{1}}=i\cos\left(\phi\right)\hat{1}+\sin\left(\phi\right)\hat{\sigma}_{1}=\left(\begin{array}{cc}
i\cos\left(\phi\right) & \sin\left(\phi\right)\\
\sin\left(\phi\right) & i\cos\left(\phi\right)
\end{array}\right)
\end{equation}
the action of the operator $\hat{L}_{1}$ on the basis states is given
by:

\begin{equation}
\hat{L}_{1}\left|s_{1}s_{2}0\right\rangle =\sin\left(\phi\right)\left|s_{1}s_{2}0\right\rangle +i\cos\left(\phi\right)\left|1-s_{1}s_{2}0\right\rangle 
\end{equation}
\begin{equation}
\hat{L}_{1}\left|s_{1}s_{2}1\right\rangle =i\cos\left(\phi\right)\left|s_{1}s_{2}1\right\rangle +\sin\left(\phi\right)\left|1-s_{1}s_{2}1\right\rangle 
\end{equation}

The operator $\hat{L}_{1}$ can be considered in terms of a quantum
regime switching model, such that if the market state neuron $n_{3}(k)$
is not firing, then, $\sin\left(\phi\right)$ is the amplitude associated
to the alternative where the neuron $n_{1}(k)$ does not change state,
while $i\cos\left(\phi\right)$ is the amplitude associated to the
alternative where the neuron $n_{1}(k)$ changes state, on the other
hand, if the neuron $n_{3}(k)$ is firing the role of the amplitudes
flip: $i\cos\left(\phi\right)$ is associated with the alternative
where the neuron $n_{1}(k)$ does not change state and $\sin\left(\phi\right)$
is the amplitude associated with the alternative where the neuron
$n_{1}(k)$ changes state.

Before considering the financial implications of this dynamics, it
is necessary to address the rest of the network, because the final
dynamics and its financial implications can only be fully addressed
at the end of the cycle. As we will see, the end result will be a
quantum computation-based selection process of adaptive rules regarding
market expectations and the processing of how financial news may support
trading decisions affecting market polarization and market volume.

Proceeding, then, with the neural links, the second neuron to be activated
is $n_{2}(k)$, which receives an input from the two neurons $n_{1}(k)$
and $n_{3}(k)$, this neuron will play a key role in the selection
of adaptive rules regarding the relation between trading profiles
and financial news, a point that we will return to when the final
neural network state transition is analyzed. Following the quantum
circuit framework, the second neuron is transformed conditionally
on the states of the two neurons $n_{1}(k)$ and $n_{3}(k)$, in accordance
with the neural links $n_{1}(k)\rightarrow n_{2}(k)\leftarrow n_{3}(k)$,
the corresponding neural links operator is given by:
\begin{equation}
\hat{L}_{2}=\sum_{s,s'\in\mathbb{A}_{2}}\left|s\right\rangle \left\langle s\right|\otimes e^{-\frac{i}{\hbar}\triangle t\hat{H}_{ss'}}\otimes\left|s'\right\rangle \left\langle s'\right|
\end{equation}
When the input neurons have synchronized firing patterns, the rotation
and phase transformation angles are set to:
\begin{equation}
\frac{\theta(00)\triangle t}{2}=\frac{\theta(11)\triangle t}{2}=0
\end{equation}
\begin{equation}
\frac{\omega(00)\triangle t}{2}=\frac{\omega(11)\triangle t}{2}=0
\end{equation}
which means that the operators reduce to:
\begin{equation}
e^{-\frac{i}{\hbar}\triangle t\hat{H}_{00}}=e^{-\frac{i}{\hbar}\triangle t\hat{H}_{11}}=\hat{1}
\end{equation}
that is, the second neuron remains in the same state when the input
neurons ($n_{1}(k)$ and $n_{3}(k)$) exhibit a synchronized firing
pattern (no rotation nor phase transformation takes place). When the
input neurons do not exhibit a synchronized firing pattern, the rotation
and phase transformation is set by the following parameters: 
\begin{equation}
\frac{\theta(01)\triangle t}{2}=\frac{\theta(10)\triangle t}{2}=\frac{\pi}{2}
\end{equation}
\begin{equation}
\frac{\omega(01)\triangle t}{2}=\frac{\omega(10)\triangle t}{2}=\frac{\pi}{2}
\end{equation}
\begin{equation}
\mathbf{u}(01)=\mathbf{u}(10)=(1,0,0)
\end{equation}
which leads to:
\begin{equation}
e^{-\frac{i}{\hbar}\triangle t\hat{H}_{01}}=e^{-\frac{i}{\hbar}\triangle t\hat{H}_{10}}=\hat{\sigma}_{1}
\end{equation}
thus, the action of $\hat{L}_{2}$ on each basis state is such that:
\begin{equation}
\hat{L}_{2}\left|ss_{2}s\right\rangle =\left|ss_{2}s\right\rangle 
\end{equation}
\begin{equation}
\hat{L}_{2}\left|ss_{2}1-s\right\rangle =\left|s1-s_{2}1-s\right\rangle 
\end{equation}
that is, the neuron $n_{2}(k)$ does not change state when the two
neurons $n_{1}(k)$ and $n_{3}(k)$ have the same firing pattern,
and flips state when the two neurons have differing firing patterns
(this is equivalent to a controlled negation quantum circuit).

Now, to close the cycle, and before addressing the final dynamics
and its financial interpretation, we have to address, first, the third
link $n_{2}(k)\rightarrow n_{3}(k)$. In this case, we also introduce
a controlled-negation circuit, so that the corresponding operator
is:
\begin{equation}
\hat{L}_{3}=\sum_{s\in\mathbb{A}_{2}}\left|s\right\rangle \left\langle s\right|\otimes\left|0\right\rangle \left\langle 0\right|\otimes e^{-\frac{i}{\hbar}\triangle t\hat{H}_{0}}+\sum_{s'\in\mathbb{A}_{2}}\left|s'\right\rangle \left\langle s'\right|\otimes\left|1\right\rangle \left\langle 1\right|\otimes e^{-\frac{i}{\hbar}\triangle t\hat{H}_{1}}
\end{equation}
\begin{equation}
\frac{\theta(0)\triangle t}{2}=0,\:\frac{\theta(1)\triangle t}{2}=\frac{\pi}{2}
\end{equation}
\begin{equation}
\frac{\omega(0)\triangle t}{2}=0,\:\frac{\omega(1)\triangle t}{2}=\frac{\pi}{2}
\end{equation}
\begin{equation}
\mathbf{u}(1)=(1,0,0)
\end{equation}
leading to:
\begin{equation}
e^{-\frac{i}{\hbar}\triangle t\hat{H}_{0}}=\hat{1},\:e^{-\frac{i}{\hbar}\triangle t\hat{H}_{1}}=\hat{\sigma}_{1}
\end{equation}
so that the basis states transform as:
\begin{equation}
\hat{L}_{3}\left|s_{1}0s_{3}\right\rangle =\left|s_{1}0s_{3}\right\rangle 
\end{equation}
\begin{equation}
\hat{L}_{3}\left|s_{1}1s_{3}\right\rangle =\left|s_{1}11-s_{3}\right\rangle 
\end{equation}
these equations show that the neuron $n_{3}(k)$ changes state when
the second neuron is firing and does not change state when the second
neuron is not firing. The neural network operator $\hat{L}_{Net}$
is the product of the three operators, that is:

\begin{equation}
\hat{L}_{Net}=\hat{L}_{3}\hat{L}_{2}\hat{L}_{1}
\end{equation}

The following table shows the results of the action of the neural
network operator on each basis state.

\begin{table}[H]
\begin{centering}
\begin{tabular}{|c|c|}
\hline 
Basis States & $\hat{L}_{Net}\left|\mathbf{s}\right\rangle =\hat{L}_{3}\hat{L}_{2}\hat{L}_{1}\left|\mathbf{s}\right\rangle $\tabularnewline
\hline 
\hline 
$\left|000\right\rangle $ & $\sin\left(\phi\right)\left|000\right\rangle +i\cos\left(\phi\right)\left|111\right\rangle $\tabularnewline
\hline 
$\left|001\right\rangle $ & $i\cos\left(\phi\right)\left|010\right\rangle +\sin\left(\phi\right)\left|101\right\rangle $\tabularnewline
\hline 
$\left|010\right\rangle $ & $\sin\left(\phi\right)\left|011\right\rangle +i\cos\left(\phi\right)\left|100\right\rangle $\tabularnewline
\hline 
$\left|011\right\rangle $ & $i\cos\left(\phi\right)\left|001\right\rangle +\sin\left(\phi\right)\left|110\right\rangle $\tabularnewline
\hline 
$\left|100\right\rangle $ & $i\cos\left(\phi\right)\left|000\right\rangle +\sin\left(\phi\right)\left|111\right\rangle $\tabularnewline
\hline 
$\left|101\right\rangle $ & $\sin\left(\phi\right)\left|010\right\rangle +i\cos\left(\phi\right)\left|101\right\rangle $\tabularnewline
\hline 
$\left|110\right\rangle $ & $i\cos\left(\phi\right)\left|011\right\rangle +\sin\left(\phi\right)\left|100\right\rangle $\tabularnewline
\hline 
$\left|111\right\rangle $ & $\sin\left(\phi\right)\left|001\right\rangle +i\cos\left(\phi\right)\left|110\right\rangle $\tabularnewline
\hline 
\end{tabular}
\par\end{centering}

\protect\caption{Neural network operator's action on the basis states.}
\end{table}

From a financial perspective, table 1 synthesizes two adaptive rules,
one in which the new market state for the component follows the new
market conditions underlying the corresponding component's dynamics
(neurons' $n_{1}(k)$ and $n_{3}(k)$ show a neural reinforcement
dynamics), and another in which the new market state is contrarian
with respect to the new market conditions underlying the corresponding
component's dynamics (neurons' $n_{1}(k)$ and $n_{3}(k)$ show a
neural inhibitory dynamics). These are two basic rules regarding expectation
formation from new data: the decision to follow the new data or not.

In the first case, and taking as example a volatility component, the
market is driven by an expectation of continuance of market conditions,
so that, for instance, if market conditions are favorable to a high
volatility state (neuron $n_{1}(k)$ is firing), then, the new market
state follows the market conditions and $n_{3}(k)$ fires, corresponding
to high volatility. 

On the other hand, still under the first adaptive rule, if market
conditions are unfavorable to a high volatility state (neuron $n_{1}(k)$
is not firing), then, the new market state follows the market conditions
and $n_{3}(k)$ does not fire, corresponding to low volatility.

The resulting adaptive rule corresponds, thus, to a \emph{follow the
news} rule. Likewise, if we consider, instead, the market polarization
component, the \emph{follow the news} rule means that if the new market
conditions support a bullish market sentiment, then, the market becomes
bullish and if the new market conditions support a bearish market
sentiment, then, the market becomes bearish.

The second adaptive rule is the reverse, expectations are that the
new market conditions will not hold, and the market does the opposite
from the news, expecting speculative gains. 

The first adaptive rule is implemented when the second neuron is not
firing, while the second rule is implemented when the second neuron
is firing. Thus, the firing of the second neuron is a dynamical component
that simulates a market change in its expectation and trading profile,
so that. for the neural configurations $\left\{ \left|000\right\rangle ,\left|001\right\rangle ,\left|100\right\rangle ,\left|101\right\rangle \right\} $,
the state transition for the market component's dynamics is driven
by the first adaptive rule, while, for the neural configurations $\left\{ \left|010\right\rangle ,\left|011\right\rangle ,\left|110\right\rangle ,\left|111\right\rangle \right\} $,
the market component's dynamics is driven by the second adaptive rule.

While neuron $n_{2}(k)$'s firing pattern determines the selection
of a \emph{follow the news rule}, the combination of firing patterns
of the three neurons determines the quantum amplitudes for the market
state transitions. Thus, when the neuron $n_{2}(k)$ is not firing,
if the initial market conditions are aligned with the initial market
state, then: $\sin\left(\phi\right)$ is the amplitude associated
with the alternative in which $n_{1}(k)$ and $n_{3}(k)$ transition
to a not firing state and $i\cos\left(\phi\right)$ is the amplitude
associated with the alternative in which $n_{1}(k)$ and $n_{3}(k)$
transition to a firing state.

For a market volatility component, this means that a transition to
a high volatility state, supported by market conditions, has an associated
quantum amplitude of $i\cos\left(\phi\right)$, while a market transition
to a low volatility, state supported by market conditions, has an
associated amplitude of $\sin\left(\phi\right)$. The role of these
amplitudes switches when $n_{1}(k)$ and $n_{3}(k)$ are not initially
aligned.

When the neuron $n_{2}(k)$ is firing, the transition amplitudes to
firing/non-firing states follow the same pattern as above for neuron
$n_{1}(k)$ but reverse the pattern for neuron $n_{3}(k)$ because
the new market conditions' neuron and the market state neuron transition
to a non-aligned state (the market is contrarian with respect to the
news), so that, if $n_{1}(k)$ and $n_{3}(k)$ are initially aligned,
$\sin\left(\phi\right)$ is the amplitude associated with a transition
to the state where $n_{1}(k)$ is not firing and $n_{3}(k)$ is firing,
while, if $n_{1}(k)$ and $n_{3}(k)$ are not initially aligned, the
amplitude associated with such a transition is $i\cos\left(\phi\right)$.
The roles of the amplitudes, thus, depend upon the way in which the
market adapts to new information and the previous configuration of
market conditions and market state.

As expected, the market conditions and the market state neurons are
always entangled, which means that, in each case, the market state
effectively becomes like a measurement apparatus of the market conditions,
the entanglement profile can, however, be aligned (\emph{follow the
news rule}, based on an expectation of sustainability of the new market
conditions) or non-aligned (\emph{contrarian rule}, based on the expectation
of reversal of the new market conditions).

Thus, in the model, the quantum neural dynamics models a market that
processes the information on the market conditions implementing a
standard quantum measurement, but the profile of that quantum measurement
depends upon the expectations regarding the news (leading to different
entanglement profiles).

The final dynamics for the market component results from the iterative
application of the operator $\hat{L}_{Net}$ for each trading round,
leading to state transition between the adaptive rules and, thus,
between the market states. Considering a sequence of neural states
for the market component's associated neural network $\left|\psi(k,t)\right\rangle $,
the state transition resulting from the dynamical rule is given by:

\begin{equation}
\left|\psi(k,t)\right\rangle =\hat{L}_{Net}\left|\psi(k,t-\triangle t)\right\rangle 
\end{equation}
leads to the following update rule for the quantum amplitudes $\psi_{k}$
(as per the general Eq.(33)):
\begin{equation}
\begin{aligned}\psi_{k}(\mathbf{s},t)=\sum_{\mathbf{s'}}\left\langle \mathbf{s}\left|\hat{L}_{Net}\right|\mathbf{s'}\right\rangle \left\langle \mathbf{s'}|\psi(k,t-\triangle t)\right\rangle =\\
=\sum_{\mathbf{s}'}L_{Net}(\mathbf{s},\mathbf{s'})\psi_{k}(\mathbf{s'},t-\triangle t)
\end{aligned}
\end{equation}
using Table 1's results, in conjunction with this last eqution, we
obtain the following transition table for the quantum amplitudes:

\begin{table}[H]
\begin{centering}
\begin{tabular}{|c|}
\hline 
New Amplitudes\tabularnewline
\hline 
\hline 
$\psi_{k}(000,t)=\sin\left(\phi\right)\psi_{k}(000,t-\triangle t)+i\cos\left(\phi\right)\psi_{k}(100,t-\triangle t)$\tabularnewline
\hline 
$\psi_{k}(001,t)=i\cos\left(\phi\right)\psi_{k}(011,t-\triangle t)+\sin\left(\phi\right)\psi_{k}(111,t-\triangle t)$\tabularnewline
\hline 
$\psi_{k}(010,t)=i\cos\left(\phi\right)\psi_{k}(001,t-\triangle t)+\sin\left(\phi\right)\psi_{k}(101,t-\triangle t)$\tabularnewline
\hline 
$\psi_{k}(011,t)=\sin\left(\phi\right)\psi_{k}(010,t-\triangle t)+i\cos\left(\phi\right)\psi_{k}(110,t-\triangle t)$\tabularnewline
\hline 
$\psi_{k}(100,t)=i\cos\left(\phi\right)\psi_{k}(010,t-\triangle t)+\sin\left(\phi\right)\psi_{k}(110,t-\triangle t)$\tabularnewline
\hline 
$\psi_{k}(101,t)=\sin\left(\phi\right)\psi_{k}(001,t-\triangle t)+i\cos\left(\phi\right)\psi_{k}(101,t-\triangle t)$\tabularnewline
\hline 
$\psi_{k}(110,t)=\sin\left(\phi\right)\psi_{k}(011,t-\triangle t)+i\cos\left(\phi\right)\psi_{k}(111,t-\triangle t)$\tabularnewline
\hline 
$\psi_{k}(111,t)=i\cos\left(\phi\right)\psi_{k}(000,t-\triangle t)+\sin\left(\phi\right)\psi_{k}(100,t-\triangle t)$\tabularnewline
\hline 
\end{tabular}
\par\end{centering}

\protect\caption{Update of the quantum amplitudes for a single market component.}
\end{table}

Taking into account this general neural dynamics for each component
we can now piece it all together to address the market state and resulting
financial dynamics.

\subsubsection{Financial Market Dynamics}

To address the full market dynamics we need to recover the QNA. For
each trading round, the quantum state associated with the market dynamics
is given by the QNA state defined as the tensor product of the lattice
site's neural networks' states, that is, by the tensor product of
each component's neural network state: 
\begin{equation}
\left|\psi(t)\right\rangle =\bigotimes_{k=1}^{N+1}\left|\psi(k,t)\right\rangle =\Psi(\mathbf{s}_{1},\mathbf{s}_{2},...,\mathbf{s}_{N+1},t)\left|\mathbf{s}_{1},\mathbf{s}_{2},...,\mathbf{s}_{N+1}\right\rangle 
\end{equation}
where the quantum amplitudes $\Psi(\mathbf{s}_{1},\mathbf{s}_{2},...,\mathbf{s}_{N+1},t)$
are given by:
\begin{equation}
\Psi(\mathbf{s}_{1},\mathbf{s}_{2},...,\mathbf{s}_{N+1},t)=\prod_{k=1}^{N+1}\psi_{k}(\mathbf{s}_{k},t)
\end{equation}
with $\psi_{k}$ being the amplitudes associated with the lattice
site $k$'s neural network.

For the $N$ volatility components we can introduce a corresponding
volatility operator on the QNA Hilbert space:
\begin{equation}
\hat{O}_{k}=\hat{1}^{\otimes k-1}\otimes\hat{O}_{V}\otimes\hat{1}^{\otimes N+1-k}
\end{equation}
with $k=1,...,N$, where, as before, $\hat{1}^{\otimes m}$ denotes
$m$-tensor product of the unit operator on $\mathcal{H}_{2}$ and
$\hat{O}_{V}$ is the volatility operator defined in Eqs.(40) and
(41). Similarly, for the market polarization operator, we write:
\begin{equation}
\hat{O}_{N+1}=\hat{1}^{\otimes N}\otimes\hat{O}_{P}
\end{equation}
where $\hat{O}_{P}$ is the market polarization operator defined in
Eqs.(42) and (43). In this way, the returns' dynamical variable defined
in Eq.(8) is replaced, in the Quantum Econophysics setting, by a quantum
operator on the QNA Hilbert space defined as:
\begin{equation}
\hat{R}=\frac{1}{\lambda}\prod_{k=1}^{N+1}\hat{O}_{k}
\end{equation}
For each basis state of the QNA Hilbert space, the returns operator
has an eigenvalue given by the corresponding financial market returns:
\begin{equation}
\hat{R}\left|\mathbf{s}_{1},\mathbf{s}_{2},...,\mathbf{s}_{N+1}\right\rangle =R(\mathbf{s}_{1},\mathbf{s}_{2},...,\mathbf{s}_{N+1})\left|\mathbf{s}_{1},\mathbf{s}_{2},...,\mathbf{s}_{N+1}\right\rangle 
\end{equation}
with the eigenvalues $R(\mathbf{s}_{1},\mathbf{s}_{2},...,\mathbf{s}_{N+1})$
given by:
\begin{equation}
R(\mathbf{s}_{1},\mathbf{s}_{2},...,\mathbf{s}_{N+1})=\frac{1}{\lambda}\prod_{k=1}^{N}v_{k}(\mathbf{s}_{k})\cdot\sigma_{N+1}(\mathbf{s}_{N+1})
\end{equation}
where $v_{k}(\mathbf{s}_{k})=v_{0}$ if the binary string $\mathbf{s}_{k}\in\mathbb{A}_{2}^{3}$
ends in $0$ ($n_{3}(k)$ is not firing) and $v_{k}(\mathbf{s}_{k})=v_{1}$
if the binary string $\mathbf{s}_{k}$ ends in $1$ ($n_{3}(k)$ is
firing), similarly $\sigma_{N+1}(\mathbf{s}_{N+1})=-1$ if $\mathbf{s}_{N+1}\in\mathbb{A}_{2}^{3}$
ends in $0$ ($n_{3}(N+1)$ is not firing) and $\sigma_{N+1}(\mathbf{s}_{N+1})=1$
if $\mathbf{s}_{N+1}$ ends in $1$ ($n_{3}(N+1)$ is firing).

The dynamical rule that comes from the neural networks' quantum computation
leads to the market state transition for each trading round:
\begin{equation}
\left|\psi(t)\right\rangle =\bigotimes_{k=1}^{N+1}\hat{L}_{Net}\left|\psi(k,t-\triangle t)\right\rangle 
\end{equation}
leading to the expected value for the returns:
\begin{equation}
\left\langle \hat{R}\right\rangle _{t}=\sum_{\mathbf{s}_{1},\mathbf{s}_{2},...,\mathbf{s}_{N+1}}R(\mathbf{s}_{1},\mathbf{s}_{2},...,\mathbf{s}_{N+1})\left|\Psi(\mathbf{s}_{1},\mathbf{s}_{2},...,\mathbf{s}_{N+1},t)\right|^{2}
\end{equation}
so that the market tends to the alternative $R(\mathbf{s}_{1},\mathbf{s}_{2},...,\mathbf{s}_{N+1})$
with an associated probability of $\left|\Psi(\mathbf{s}_{1},\mathbf{s}_{2},...,\mathbf{s}_{N+1},t)\right|^{2}$.

The following figure shows a market simulation on Python 3.4. In the
simulations, the initial state for each component is taken from a
randomly chosen $\textrm{U}(2)$ gate applied to each neuron with
uniform probability over $\textrm{U}(2)$ . The figure shows the markers
of financial turbulence in the returns, including volatility bursts
and jumps.

\begin{figure}[H]
\begin{centering}
\includegraphics[scale=0.5]{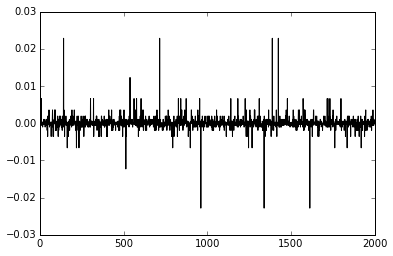}
\par\end{centering}

\protect\caption{Simulation of the financial returns for $\sin^{2}\phi=0.6$, $v_{0}=0.7$,
$\lambda=1000$, 20 components (19 volatility components plus 1 polarization
component). The figure shows 2000 data points of a 2100 data points
simulation with the first 100 points removed for transients.}
\end{figure}

The main parameters that determine the market profile with regards
to turbulence is $v_{0}$ and the number of components, the turbulence
profile does not change much with respect to the rotation angle $\phi$.
Indeed, as shown in the following table, the estimated kurtosis\footnote{The Fisher kurtosis is used in the statistical analysis of the model's
outputs.} for different simulations with 20 components tends to decrease as
$v_{0}$ rises. For $v_{0}=0.9$ we no longer find excess kurtosis,
the turbulence markers being lost. This approach to low turbulence
is progressive as $v_{0}$ is raised from $0.8$ to $0.9$, such that
that the price jumps tend to become less severe and less frequent,
and the volatility bursts tend to disappear, as shown in Figure 2,
in which $v_{0}=0.9$ with the rest of the parameters used in Figure
1's simulation being left unchanged.

\begin{table}[H]
\begin{centering}
{\small{}}%
\begin{tabular}{|c|c|c|c|}
\hline 
 & {\small{}$\sin^{2}\phi=0.4$} & {\small{}$\sin^{2}\phi=0.5$} & {\small{}$\sin^{2}\phi=0.6$}\tabularnewline
\hline 
\hline 
{\small{}$v_{0}=0.4$} & {\small{}585.4546} & {\small{}1336.6726} & {\small{}1923.3159}\tabularnewline
\hline 
{\small{}$v_{0}=0.5$} & {\small{}778.0387} & {\small{}1876.3810} & {\small{}783.0852}\tabularnewline
\hline 
{\small{}$v_{0}=0.6$} & {\small{}1015.5296} & {\small{}473.4505} & {\small{}383.9775}\tabularnewline
\hline 
{\small{}$v_{0}=0.7$} & {\small{}77.6054} & {\small{}49.5857} & {\small{}56.8335}\tabularnewline
\hline 
{\small{}$v_{0}=0.8$} & {\small{}6.8827} & {\small{}20.7037} & {\small{}5.6217}\tabularnewline
\hline 
{\small{}$v_{0}=0.9$} & {\small{}-0.8538} & {\small{}-1.2335} & {\small{}-0.9277}\tabularnewline
\hline 
\end{tabular}
\par\end{centering}{\small \par}

\protect\caption{Kurtosis values for different values of $\sin^{2}\phi$ and $v_{0}$.
The other parameters are: $\lambda=1000$, 20 components (19 volatility
components plus 1 polarization component), the Kurtosis coefficient
was calculated on 5000 sample data points of a 5100 data points simulation
with the first 100 data points removed for transients.}
\end{table}

\begin{figure}[H]
\begin{centering}
\includegraphics[scale=0.5]{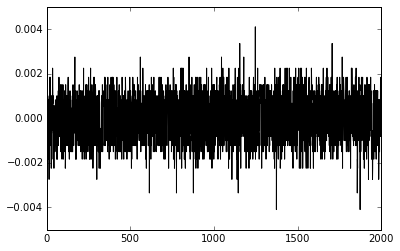}
\par\end{centering}

\protect\caption{Simulation of financial returns for $\sin^{2}\phi=0.6$, $v_{0}=0.9$,
$\lambda=1000$, 20 components (19 volatility components plus 1 polarization
component). The figure shows 2000 data points of a 2100 data points
simulation with the first 100 points removed for transients.}
\end{figure}

The model, thus, captures different market profiles: as the parameter
$v_{0}$ increases from $0.8$ to $0.9$ the simulations tend to approach
a lower tail risk dynamics, with a greater approximation to the classical
Gaussian returns' probability model ocurring for $v_{0}$ near $0.87$,
the following table shows this approximation with the kurtosis values
and Jarque-Bera test for normality, as the value of $v_{0}$ is increased.

\begin{table}[H]
\begin{centering}
{\small{}}%
\begin{tabular}{|c|c|c|c|}
\hline 
 & {\small{}Kurtosis} & {\small{}JB Statistic} & {\small{}p-value}\tabularnewline
\hline 
\hline 
{\small{}$v_{0}=0.85$} & {\small{}1.2911} & {\small{}353.7757} & {\small{}0.0}\tabularnewline
\hline 
{\small{}$v_{0}=0.86$} & {\small{}0.3596} & {\small{}46.5669} & {\small{}7.7289e-11}\tabularnewline
\hline 
{\small{}$v_{0}=0.87$} & {\small{}0.1746} & {\small{}6.3179} & {\small{}0.0425}\tabularnewline
\hline 
{\small{}$v_{0}=0.88$} & {\small{}-0.0160} & {\small{}76.5797} & {\small{}0.0}\tabularnewline
\hline 
{\small{}$v_{0}=0.89$} & {\small{}-0.6790} & {\small{}106.5589} & {\small{}0.0}\tabularnewline
\hline 
{\small{}$v_{0}=0.9$} & {\small{}-0.8223} & {\small{}143.5983} & {\small{}0.0}\tabularnewline
\hline 
\end{tabular}
\par\end{centering}{\small \par}

\protect\caption{Kurtosis values and Jarque-Bera test of normality for different values
$v_{0}$. The other parameters are: $\sin^{2}\phi=0.6$, $\lambda=1000$,
20 components (19 volatility components plus 1 polarization component),
the Kurtosis coefficient was calculated on 5000 sample data points
of a 5100 data points simulation with the first 100 data points removed
for transients.}
\end{table}

As shown in table 4 for every value of $v_{0}$ the Jarque-Bera's
null hypothesis is rejected at 1\% significance level except for $v_{0}=0.87$.
It is important to stress however, that although simulated returns
distribution can approximate the Gaussian distribution, this approximation
is not robust, different simulations for the same parameters may show
deviations from the Gaussian distribution.

The following table shows examples of simulations for different values
of the rotation angle $\phi$, with $v_{0}=0.87$, the null hypothesis
of Jarque-Bera's test is not reject, at a 1\% significance, for $\sin^{2}\phi=0.1,\:0.3,\:0.6$,
with $\sin^{2}\phi=0.1,\:0.3$ as the only cases in which it is not
rejected for 5\% significance, and $\sin^{2}\phi=0.3$ as the only
case in which it is not rejected also for a 10\% significance.

\begin{table}[H]
\begin{centering}
{\small{}}%
\begin{tabular}{|c|c|c|c|}
\hline 
$\sin^{2}\phi$ & {\small{}Kurtosis} & {\small{}JB Statistic} & {\small{}p-value}\tabularnewline
\hline 
\hline 
{\small{}$0.1$} & {\small{}0.1285} & {\small{}4.8697} & {\small{}0.0876}\tabularnewline
\hline 
{\small{}$0.2$} & {\small{}0.6385} & {\small{}84.4538} & {\small{}0.0}\tabularnewline
\hline 
{\small{}$0.3$} & {\small{}0.0947} & {\small{}3.8794} & {\small{}0.1437}\tabularnewline
\hline 
{\small{}$0.4$} & {\small{}0.3169} & {\small{}35.1954} & {\small{}2.2773e-08}\tabularnewline
\hline 
{\small{}$0.5$} & {\small{}0.9213} & {\small{}565.5135} & {\small{}0.0}\tabularnewline
\hline 
{\small{}$0.6$} & {\small{}0.1746} & {\small{}6.3179} & {\small{}0.0425}\tabularnewline
\hline 
{\small{}$0.7$} & {\small{}0.0841} & {\small{}29.6076} & {\small{}3.7221e-07}\tabularnewline
\hline 
{\small{}$0.8$} & {\small{}0.4975} & {\small{}65.859} & {\small{}4.9960e-15}\tabularnewline
\hline 
{\small{}$0.9$} & {\small{}0.1775} & {\small{}31.3417} & {\small{}1.5640e-07}\tabularnewline
\hline 
\end{tabular}
\par\end{centering}{\small \par}

\protect\caption{Kurtosis values for different values $\phi$. The other parameters
are: $v_{0}=0.87$ $\lambda=1000$, 20 components (19 volatility components
plus 1 polarization component), the Kurtosis coefficient was calculated
on 5000 sample data points of a 5100 data points simulation with the
first 100 data points removed for transients.}
\end{table}

These results may, however, depend, as stated previously, upon the
simulation, other simulations may show the null hypothesis being rejected
for the same parameters, which means that the Gaussian distribution
depends upon the sample path and is not a dynamically fixed probability
law that can be assumed to hold indefinitely.

The general tail risk pattern, on the other hand, is more robust than
the Gaussian approximation, in the sense that as $v_{0}$ approaches
$0.9$ and for $v_{0}\geq0.9$, the market loses the turbulence profile
with the jumps and volatility changes becoming less frequent and the
kurtosis becoming less and less leptokurtic, leading to lower tail
risk, the market returns eventually fluctuate randomly around a narrow
band.

Underlying the complex behavior of the simulated market returns, is
the probability dynamics that comes from the neural network's iterative
scheme shown in table 2. Considering Eqs.(35) to (37) and combining
with table 2's results we get, in this case, sixteen nonlinear dynamical
equations of the general form:
\begin{equation}
A_{k}(\mathbf{s},t)=\left[\sqrt{A_{k}(\mathbf{s'},t-\triangle t)}\sin\left(\phi\right)-\sqrt{B_{k}(\mathbf{s''},t-\triangle t)}\cos\left(\phi\right)\right]^{2}
\end{equation}
\begin{equation}
B_{k}(\mathbf{s},t)=\left[\sqrt{B_{k}(\mathbf{s'},t-\triangle t)}\sin\left(\phi\right)+\sqrt{A_{k}(\mathbf{s''},t-\triangle t)}\cos\left(\phi\right)\right]^{2}
\end{equation}
with $\mathbf{s},\mathbf{s'},\mathbf{s''}\in\mathbb{A}_{2}^{3}$ and
$\mathbf{s'}\neq\mathbf{s''}$, so that the probability dynamics that
come from the neural network's evolution can be addressed by a nonlinear
map with sixteen dynamical variables satisfying the normalization
rule:
\begin{equation}
\sum_{\mathbf{s}}A_{k}(\mathbf{s},t)+\sum_{\mathbf{s'}}B_{k}(\mathbf{s'},t)=1
\end{equation}
with the probability of the neural configuration $\mathbf{s}$ being
given by the sum:
\begin{equation}
Prob_{k}[\mathbf{s},t]=A_{k}(\mathbf{s},t)+B_{k}(\mathbf{s},t)
\end{equation}
so that the probability distribution for the neural configurations
is a function of a sixteen dimensional nonlinear map on a hypersphere
of unit radius (due to the normalization condition).

If we expand the squares in Eqs.(83) and (84) we get:

\begin{equation}
\begin{aligned}A_{k}(\mathbf{s},t)=A_{k}(\mathbf{s'},t-\triangle t)\sin^{2}\left(\phi\right)+B_{k}(\mathbf{s''},t-\triangle t)\cos^{2}\left(\phi\right)-\\
-\sqrt{A_{k}(\mathbf{s'},t-\triangle t)B_{k}(\mathbf{s''},t-\triangle t)}\sin\left(2\phi\right)
\end{aligned}
\end{equation}

\begin{equation}
\begin{aligned}B_{k}(\mathbf{s},t)=B_{k}(\mathbf{s'},t-\triangle t)\sin^{2}\left(\phi\right)+A_{k}(\mathbf{s''},t-\triangle t)\cos^{2}\left(\phi\right)+\\
+\sqrt{B_{k}(\mathbf{s'},t-\triangle t)A_{k}(\mathbf{s''},t-\triangle t)}\sin\left(2\phi\right)
\end{aligned}
\end{equation}
which leads to the following expansion for the probability:
\begin{equation}
\begin{aligned}Prob_{k}[\mathbf{s},t]=\\
=Prob_{k}[\mathbf{s'},t-\triangle t]\sin^{2}\left(\phi\right)+\\
+Prob_{k}[\mathbf{s''},t-\triangle t]\cos^{2}\left(\phi\right)+\\
+\sqrt{B_{k}(\mathbf{s'},t-\triangle t)A_{k}(\mathbf{s''},t-\triangle t)}\sin\left(2\phi\right)-\\
-\sqrt{A_{k}(\mathbf{s'},t-\triangle t)B_{k}(\mathbf{s''},t-\triangle t)}\sin\left(2\phi\right)
\end{aligned}
\end{equation}
the quantum interference terms (that correspond to the square root
terms multiplied by $\sin\left(2\phi\right)$ in Eq.(89)) have an
expression, at the probability level, that can be approached in terms
of a classical nonlinear dynamical system for the probabilities.

In the classical nonlinear dynamics representation, each financial
returns component's stochastic dynamics has a probability measure
that updates at each trading round with a deterministic nonlinear
update rule, this establishes the bridge between the stochastic process
and the nonlinear deterministic dynamical systems modeling of financial
dynamics: the neural network's quantum dynamics leads to a nonlinear
deterministic dynamics in the probabilities.

A question that may be raised regards the transition from the deterministic
nonlinear map to a noisy nonlinear map, from the financial perspective
this makes sense since external stochastic factors may affect the
financial system. A possible solution for this might be to allow the
rotation angle $\phi$ to change, so that instead of a fixed value
of $\phi$ we replace it by a random variable $\phi_{k}(t)$ in Eqs.(83)
and (84) so that we get a stochastic nonlinear dynamical system. The
introduction of a random $\phi_{k}(t)$ implies that we are no longer
dealing with a fixed unitary operator structure for the QuANN but,
instead, work with a quantum neural state transition with a random
component in the Hamiltonian, that is, the unitary gates of Eqs.(50)
and (51) are now stochastic unitary gates:
\begin{equation}
e^{-\frac{i}{\hbar}\triangle t\hat{H}_{0}(t)}=\left(\begin{array}{cc}
\sin\left(\phi_{k}(t)\right) & i\cos\left(\phi_{k}(t)\right)\\
i\cos\left(\phi_{k}(t)\right) & \sin\left(\phi_{k}(t)\right)
\end{array}\right)
\end{equation}
\begin{equation}
e^{-\frac{i}{\hbar}\triangle t\hat{H}_{1}(t)}=\left(\begin{array}{cc}
i\cos\left(\phi_{k}(t)\right) & \sin\left(\phi_{k}(t)\right)\\
\sin\left(\phi_{k}(t)\right) & i\cos\left(\phi_{k}(t)\right)
\end{array}\right)
\end{equation}
Thus, a stochastic nonlinear map is induced by the quantum noisy gates
in the QuANN's state transition rule, coming from a stochastic Hamiltonian.
The following figure shows the simulation results for
\begin{equation}
\phi_{k}(t)=\arcsin\left(\frac{1}{\sqrt{1+e^{-2\beta z_{k}(t)}}}\right)
\end{equation}
with $z_{k}(t)\sim N(0,1)$, which leads to: 
\begin{equation}
\sin^{2}\left(\phi_{k}(t)\right)=\frac{1}{1+e^{-2\beta z_{k}(t)}}
\end{equation}
\begin{equation}
\cos^{2}\left(\phi_{k}(t)\right)=\frac{e^{-2\beta z_{k}(t)}}{1+e^{-2\beta z_{k}(t)}}
\end{equation}
the logistic function present in Eqs.(93) and (94) is also widely
used in classical ANNs for the activation probability and leaves room
for expansion of connections to Statistical Mechanics (Müller and
Strickland, 1995). If we replace in Eq.(89) we get the nonlinear stochastic
equations for the probabilities:
\begin{equation}
\begin{aligned}Prob_{k}[\mathbf{s},t]=\\
=\frac{Prob_{k}[\mathbf{s'},t-\triangle t]}{1+e^{-2\beta z_{k}(t)}}\\
+\frac{Prob_{k}[\mathbf{s''},t-\triangle t]e^{-2\beta z_{k}(t)}}{1+e^{-2\beta z_{k}(t)}}+\\
2\sqrt{B_{k}(\mathbf{s'},t-\triangle t)A_{k}(\mathbf{s''},t-\triangle t)}\frac{e^{-\beta z_{k}(t)}}{1+e^{-2\beta z_{k}(t)}}-\\
-2\sqrt{A_{k}(\mathbf{s'},t-\triangle t)B_{k}(\mathbf{s''},t-\triangle t)}\frac{e^{-\beta z_{k}(t)}}{1+e^{-2\beta z_{k}(t)}}
\end{aligned}
\end{equation}

\begin{figure}[H]
\begin{centering}
\includegraphics[scale=0.5]{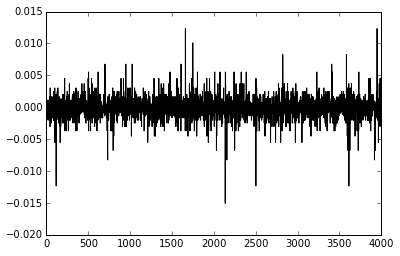}
\par\end{centering}

\protect\caption{Simulation of financial returns for noisy gates with $\beta=2.0$,
$v_{0}=0.9$, $\lambda=1000$, 80 components (79 volatility components
plus 1 polarization component). The figure shows 4000 data points
of a 4100 data points simulation with the first 100 points removed
for transients.}
\end{figure}

The figure 3 shows the occurrence of price jumps and clustering volatility,
the turbulence in this case is linked to the high number of components
(rather than to the noisy gates). Indeed, as the following two tables
show, the noisy gates do not have a strong effect on the transition
from leptokurtic to platikurtic distributions, both for low and high
values of $\beta$, it is the number of components that have a stronger
impact on market profile, as seen in table 7 for the case of $v_{0}=0.88$
which for the simulation with $\beta=2$ was close enough to the Gaussian
distribution for the non-rejection of the null hypothesis of the Jarque-Bera
test at a 10\% significance level.

\begin{table}[H]
\begin{raggedright}
{\small{}}%
\begin{tabular}{|c|c|c|c|c|c|}
\hline 
{\small{}$\beta=0.01$} & {\small{}Kurtosis} & {\small{}JB p-value} & {\small{}$\beta=2$} & {\small{}Kurtosis} & {\small{}JB p-value}\tabularnewline
\hline 
\hline 
{\small{}$v_{0}=0.86$} & {\small{}0.9171} & {\small{}0.0} & {\small{}$v_{0}=0.86$} & {\small{}0.7097} & {\small{}0.0}\tabularnewline
\hline 
{\small{}$v_{0}=0.87$} & {\small{}0.0345} & {\small{}1.0518e-07} & {\small{}$v_{0}=0.87$} & {\small{}0.1786} & {\small{}0.0018}\tabularnewline
\hline 
{\small{}$v_{0}=0.88$} & {\small{}-0.3647} & {\small{}2.8422e-13} & {\small{}$v_{0}=0.88$} & {\small{}-0.0856} & {\small{}0.3119}\tabularnewline
\hline 
{\small{}$v_{0}=0.89$} & {\small{}-0.7142} & {\small{}0.0} & {\small{}$v_{0}=0.89$} & {\small{}-0.6205} & {\small{}0.0}\tabularnewline
\hline 
{\small{}$v_{0}=0.9$} & {\small{}-0.6810} & {\small{}0.0} & {\small{}$v_{0}=0.9$} & {\small{}-0.8840} & {\small{}0.0}\tabularnewline
\hline 
\end{tabular}
\par\end{raggedright}{\small \par}

\protect\caption{Kurtosis values and Jarque-Bera test of normality p-values for different
simulations with variying $v_{0}$ and noisy unitary gates. The other
parameters are: $\beta=0.01$ (left table) and $\beta=2$ (right table),
$\lambda=1000$, 20 components (19 volatility components plus 1 polarization
component), the kurtosis coefficient was calculated on 5000 sample
data points of a 5100 data points simulation with the first 100 data
points removed for transients.}
\end{table}

\begin{table}[H]
\begin{raggedright}
{\small{}}%
\begin{tabular}{|c|c|c|c|c|c|}
\hline 
{\small{}$\beta=0.01$} & {\small{}Kurtosis} & {\small{}p-value} & {\small{}$\beta=2$} & {\small{}Kurtosis} & {\small{}p-value}\tabularnewline
\hline 
\hline 
{\small{}$N+1=10$} & {\small{}-1.3408} & {\small{}0.0} & {\small{}$N+1=10$} & {\small{}-1.3752} & {\small{}0.0}\tabularnewline
\hline 
{\small{}$N+1=20$} & {\small{}-0.3647} & {\small{}2.8422e-13} & {\small{}$N+1=20$} & {\small{}-0.0856} & {\small{}0.3119}\tabularnewline
\hline 
{\small{}$N+1=30$} & {\small{}1.6227} & {\small{}0.0} & {\small{}$N+1=30$} & {\small{}2.2008} & {\small{}0.0}\tabularnewline
\hline 
{\small{}$N+1=40$} & {\small{}2.5130} & {\small{}0.0} & {\small{}$N+1=40$} & {\small{}3.4597} & {\small{}0.0}\tabularnewline
\hline 
{\small{}$N+1=50$} & {\small{}10.1352} & {\small{}0.0} & {\small{}$N+1=50$} & {\small{}15.9802} & {\small{}0.0}\tabularnewline
\hline 
\end{tabular}
\par\end{raggedright}{\small \par}

\protect\caption{Kurtosis values and Jarque-Bera test of normality for different values
of the number of components ($N+1$) and noisy unitary gates. The
other parameters are: $v_{0}=0.88$ $\beta=0.01$ (left table) and
$\beta=2$ (right table), $\lambda=1000$, the kurtosis coefficient
was calculated on 5000 sample data points of a 5100 data points simulation
with the first 100 data points removed for transients.}
\end{table}

Indeed, the number of components shows a strong effect, as can be
seen in figure 3, which uses $v_{0}=0.9$ and in table 7, that shows
the transition from platikurtic to leptokurtic for large values of
the components, for both a low and a high value of $\beta$.

\section{Finance, Nonlinear Stochastic Dynamics and Quantum Artificial Intelligence}

From the early onset of development of Econophysics, some form of
nonlinear stochastic dynamics has been considered to be present in
financial market dynamics. A major example being Vaga's work that
addressed explicitly different probability distributions corresponding
to different (classical) Hamiltonian conditions (Vaga, 1990). The
major point that markets make transitions between different regimes
and different probability distributions was key to Vaga's market theory.
On the other hand, the multifractal multiplicative cascades (Mandelbrot
et al., 1997) introduced multiplicative stochastic processes as sources
of market turbulence.

While a division line is drawn in regards to nonlinear deterministic
processes versus nonlinear stochastic processes, the possible combination
of both might provide an intermediate approach, combining adaptive
market dynamics and stochastic factors affecting market behavior.

As the previous section model shows, when recurrent QuANNs are applied
to financial modeling, the nonlinear deterministic dynamics and the
nonlinear stochastic processes result directly from the quantum computational
structure, in the sense that: while the iterative computation of a
QuANN results from the linear conditional unitary state transition,
the corresponding probabilities, due to the square modulus rule for
addressing the probabilities associated to different neural firing
patterns, leads to a nonlinear update rule for the probabilities themselves,
which means that the market behavior will show an interference effect
at the probability level expressable in terms of a classical nonlinear
map, thus, while the system follows a stochastic dynamics, the probabilities
are updated nonlinearly.

This is a direct consequence of Quantum Cognitive Science that comes
from human decision analysis, which shows that the nonlinear update
in probabilities, leading to non-additive decision weights may be
computationally approached from linear unitary quantum computation
on an appropriate Hilbert space. Stochastic factors in the nonlinear
update of probabilities can also be introduced through unitary noise
in the neural network's computation through stochastic Hamiltonians.

Although QuAI and QuANN theory are still on their early stages, they
provide a bridge between major lines of research on financial dynamics
and risk modeling including: nonlinear deterministic and stochastic
dynamics applied to financial modeling, Cognitive Science and computational
foundations of Financial Theory. Future research on QuANNs dynamics
may thus serve as a relevant tool to link different approaches that
characterized the different lines of research on Econophysics-based
Finance.\foreignlanguage{american}{\bibliographystyle{jox}
\bibliography{Mine-not-blind,Others}
}

Anderson, Philip W., Kenneth J. Arrow and David Pines (Eds.), 1988,
The Economy as an Evolving Complex System, Perseus Books, USA.

Arthur, Brian W., Steven N. Durlauf and David Lane (Eds.), 1997, The
Economy as an Evolving Complex System II, Westview Press, USA.

Baaquie, Belal E., L.C. Kwek and M. Srikant, 2000, Simulation of Stochastic
Volatility using Path Integration: Smiles and Frowns, arXiv:cond-mat/0008327v1.

Baaquie, Belal E. and Srikant Marakani, 2001, Empirical investigation
of a quantum field theory of forward rates, arXiv:cond-mat/0106317v2
{[}cond-mat.stat-mech{]}.

Baaquie, Belal E., 2004, Quantum Finance: Path Integrals and Hamiltonians
for Options and Interest Rates, Cambridge University Press, USA.

Baaquie, Belal E. and T. Pan, 2011, Simulation of coupon bond European
and barrier options in quantum finance, Physica A, 390, 263-289.

Behrman EC., Niemel J, Steck JE, Skinner SR., 1996, A Quantum Dot
Neural Network, In: T. Toffoli T and M. Biafore, Proceedings of the
4th Workshop on Physics of Computation, Elsevier Science Publishers,
Amsterdam, 22-24.

Bohm, David, 1984, Causaity and Chance in Modern Physics, (1997 Reprinted
Version) Routledge \& Kegan Paul, Eastbourne.

Bohm, David and Basil J. Hiley, 1993, The undivided universe - An
ontological interpretation of quantum theory, Routledge, London.

Bransden, B.H. and Joachain, C.J., 2000, Quantum Mechanics, Prentice
Hall, England.

Bruce, Colin, 2004, Schrödinger's Rabbits - the many worlds of quantum,
Joseph Henry Press, Washington DC.

Brunn, Charlotte (Ed.), 2006, Advances in Artificial Economics - The
Economy as a Complex Dynamic System, Springer, Berlin.

Busemeyer, Jerome R. and Riccardo Franco, 2010, What is The Evidence
for Quantum Like Interference Effects in Human Judgments and Decision
Behavior?, NeuroQuantology, vol.8, No.4, DOI: 10.14704/nq.2010.8.4.350.

Busemeyer, Jerome R. and Peter D. Bruza, 2012, Quantum models of cognition
and decision, Cambridge University Press, Cambridge.

Busemeyer, Jerome R. and Zheng Wang, 2014, Quantum Cognition: Key
Issues and Discussion, Topics in Cognitive Science 6: 43-46.

Calvet, Laurent and A. Fisher, 2002, Multifractality in asset returns:
Theory and evidence, Review of Economics and Statistics, Vol.84, 3,
August, 381-406.

Calvet, Laurent and A. Fisher, 2004, Regime-switching and the estimation
of multifractal processes. Journal of Financial Econometrics , 2:
44-83.

Choustova, Olga, 2007a, Quantum modeling of nonlinear dynamics of
stock prices: Bohmian approach, Theoretical and Mathematical Physics,
152(2): 1213-1222.

Choustova, Olga, 2007b, Toward quantum-like modeling of financial
processes, J. Phys.: Conf. Ser. 70 01 2006.

Chrisley R., 1995, Quantum learning, In: Pylkkänen P and Pylkkö P
(eds.), New directions in cognitive science: Proceedings of the international
symposium, Saariselka, 4-9 August, Lapland, Finland, Finnish Artificial
Intelligence Society, Helsinki, 77-89.

Deutsch, David, 1985, Quantum theory, the Church-Turing Principle
and the universal quantum computer, Proc R Soc Lond A, 400-497.

Deutsch, David, 1999, Quantum theory of probability and decisions,
Proc. R. Soc. Lond. A 1999 455 3129-3137.

DeWitt, Bryce S, 1970, Quantum mechanics and reality - Could the solution
to the dilemma of indeterminism be a universe in which all possible
outcomes of an experiment actually occur?, Physics Today , Vol .23
, No .9, 155-165.

Dirac, P.A.M., 1967, The Principles of Quantum Mechanics (Fourth Edition),
Claredon Press, Oxford.

Everett, Hugh, 1957, 'Relative state' formulation of quantum mechanics.
Rev. of Mod. Physics, 29 (3):454-462.

Everett, Hugh, 1973, The Theory of the Universal Wavefunction, PhD
Manuscript, In: DeWitt, R. and N. Graham (Eds.), The Many-Worlds Interpretation
of Quantum Mechanics. Princeton Series in Physics, Princeton University
Press, 3-140.

Ehrentreich, Norman, 2008, Agent-Based Modeling - The Santa Fe Institute
Artificial Stock Market Model Revisited, Springer, Berlin.

Farmer, Doyne, 2002, Market force, ecology and evolution, Industrial
and Corporate Change, 11 (5): 895-953.

Focardi, Sergio M. and Frank J. Fabozzi, 2004, The Mathematics of
Financial Modeling and Investment Management, John Wiley and Sons,
New Jersey.

Gonçalves Carlos P, 2011, Financial turbulence, business cycles and
intrinsic time in an artificial economy, Algorithmic Finance, 1(2):
141\textendash 156.

Gonçalves, Carlos P., 2013, Quantum Financial Economics - Risk and
Returns, J Syst Sci Complex, 26: 187-200.

Gonçalves, Carlos P., 2015, Quantum Cybernetics and Complex Quantum
Systems Science - A Quantum Connectionist Exploration, NeuroQuantology,
vol.13, No.1, DOI: 10.14704/nq.2015.13.1.804.

Greiner, Walter and Brendt Müller, 2001, Quantum Mechanics Symmetries,
Springer, Germany.

Haven, Emmanuel and Andrei Khrennikov, 2013, Quantum Social Science,
Cambridge University Press, New York.

Haken, H., 1977, Synergetics: An Introduction, Springer, Germany.

Iori, Giulia, 1999, Avalanche Dynamics and Trading Friction Effects
on Stock Market Returns, Int. J. Mod. Phys. C, 10, 1149.

Ilinski, Kirill, 2001, Physics of Finance - Gauge Modelling in Non-equilibrium
Pricing, John Wiley and Sons, West Sussex.

Ivancevic VG and Ivancevic TT., 2010, Quantum Neural Computation,
Springer, Dordrecht.

Kak S., 1995, Quantum Neural Computing. Advances in Imaging and Electron
Physics, vol. 94:259-313.

Kauffman, Stuart A., 1993, The Origins of Order - Self-Organization
and Selection in Evolution, Oxford University Press, New York.

Khrennikov, Andrei, 2010, Ubiquitous Quantum Structure: From Psychology
to Finance, Springer, Berlin.

Khrennikov, Andrei and Irina Basieva, 2014, Quantum Model for Psychological
Measurements: From the Projection Postulate to Interference of Mental
Observables Represented As Positive Operator Valued Measures, NeuroQuantology,
vol.12, No.3, DOI: 10.14704/nq.2014.12.3.750.

Leggett, A.J., 2002, Qubits, Cbits, Decoherence, Quantum Measurement
and Environment, In: Heiss, (Ed.), Fundamentals of Quantum Information
- Quantum Computation, Communication, Decoherence and All That, Springer,
Germany.

Lux, Thomas, Michele Marchesi, 1999, Scaling and criticality in a
stochastic multi-agent model of a financial market, Nature 397, 498-500.

Lux, Thomas, 2008, The Markov-Switching Multifractal Model of Asset
returns: GMM estimation and linear forecasting of volatility. Journal
of Business and Economic Statistics , 26:194-210.

Mandelbrot, Benoit B., A. Fisher and L. Calvet, 1997, A Multifractal
Model of Asset Returns, Cowles Foundation Discussion Papers: 1164.

Mandelbrot, Benoit, B., 1997, Fractals and Scaling in Finance, Springer,
USA.

McCulloch, W. and W. Pitts, 1943, A logical calculus of the ideas
immanent in nervous activity. Bulletin of Mathematical Biophysics,
7:115 - 133.

Menneer T and Narayanan A., 1995, Quantum-inspired Neural Networks,
technical report R329, Department of Computer Science, University
of Exeter, Exeter, United Kingdom.

Menneer T., 1998, Quantum Artificial Neural Networks, Ph. D. thesis,
The University of Exeter, UK.

Müller B, J. Reinhardt and M.T. Strickland, 1995, Neural Networks
An Introduction, Springer-Verlag, Berlin.

Nash, J., 1951, Non-Cooperative Games, The Annals of Mathematics,
Second Series,Vol. 54, No. 2, 286-295.

Nielsen, Michael and Issac Chuang, 2003, Quantum Computation and Quantum
Information, The Press Syndicate of the University of Cambridge, UK.

Piotrowski, Edward W. and J. Sladkowski, 2001, Quantum-like approach
to financial risk: quantum anthropic principle, Acta Phys.Polon.,
B32, 3873.

Piotrowski, Edward W. and J. Sladkowski, 2002, Quantum Market Games,
Physica A, Vol.312, 208-216.

Piotrowski, Edward W. and J. Sladkowski, 2008, Quantum auctions: Facts
and myths, Physica A, Vol. 387, 15, 3949-3953.

Saptsin, V. and V. Soloviev, 2009, Relativistic quantum econophysics
- new paradigms in complex systems modelling, arXiv:0907.1142v1 {[}physics.soc-ph{]}.

Saptsin, V. and V. Soloviev, 2011, Heisenberg uncertainty principle
and economic analogues of basic physical quantities, arXiv:1111.5289v1
{[}physics.gen-ph{]}.

Segal, W. and I.E. Segal, 1998, The Black\textendash Scholes pricing
formula in the quantum context. PNAS March 31, vol. 95 no. 7, 4072-4075.

Vaga, Tonis, 1990, The Coherent Market Hypothesis, Financial Analysts
Journal, 46, 6: 36-49.

Voit, Johannes, 2001, The Statistical Mechanics of Financial Markets,
Springer, New York.

Wallace, David, 2002, Quantum Probability and Decision Theory, Revisited,
arXiv:quant-ph/0211104.

Wallace, David, 2007, Quantum Probability from Subjective Likelihood:
improving on Deutsch's proof of the probability rule, Studies in the
History and Philosophy of Modern Physics 38, 311-332.

Wang, Zheng and Jerome R. Busemeyer, 2013, A quantum question order
model supported by empirical tests of an a priori and precise prediction,
Topics in Cognitive Science, 5: 689\textendash 710.

Zuo, Xingdong, 2014, Numerical Simulation of Asano-Khrennikov-Ohya
Quantum-like Decision Making Model, vol.12, No.4, DOI: 10.14704/nq.2014.12.4.778.
\end{document}